\documentclass[10pt]{article} 
\usepackage{amsfonts, amssymb, amsmath, xcolor}
\newcommand{\id}{\mbox{id}}

\newcommand{\tr}{\mbox{Tr}}

\renewcommand{\hat}{\widehat}
\newcommand{\Tesp}{{\bf Tesp}}

\newcommand{\GTesp}{\mbox{G-}\Tesp}

\newcommand{\Set}{{\bf FinSet}} 
\newcommand{\SET}{{\bf Set}_{0}}
\newcommand{\Grel}{{\bf Grel}}
\newcommand{\Rel}{{\bf FRel}}
\newcommand{\G}{{\mathfrak G}}
\newcommand{\Gr}{\frak{Gr}}
\newcommand{\Grp}{{\bf Grp}}
\newcommand{\Sinj}{{\bf FinInj}} 
\newcommand{\FinInj}{{\bf FinInj}}

\newcommand{\C}{{\mathbb C}}
\newcommand{\Hilb}{{\cal H}}
\renewcommand{\H}{{\cal H}} 
\newcommand{\co}{\text{\bf co}}

\newcommand{\red}{}
\newcommand{\blue}{}
\newcommand{\green}{}

\input xy
\xyoption{all}

\newcommand{\Hord}{{\bf \Hord}}
\newcommand{\ran}{\text{ran}}
\renewcommand{\1}{{\sf 1}}

\newcounter{thaler}
\newenvironment{mlist}
{\begin{list}{\arabic{thaler}}%
{\usecounter{thaler} \setlength{\rightmargin}{\leftmargin}
\oddsidemargin=.3in \topsep=5pt
\itemsep=5pt
\parskip=0pt
\parsep=0pt
}}{\end{list}}

\begin{document}

\begin{center}{\large \bf Symmetry and Composition in Probabilistic Theories}\footnote{\red The present version of this paper corrects a large number of typographical 
and other editing errors in the published version ({\em Electronic Notes in Theor. Comp. Sci.} {\bf 270}, 2011). 
Sections 4, 5 and 6 have also been extensively revised 
and expanded to improve readability, and include some new results. A new section 7 replaces the 
seriously flawed discussion of composite systems and the monoidality of the category $G$-$\Tesp$ in the earlier 
version. I'd like to thank Jason Morton, whose invitation to speak in the Penn State Applied Algebra and Network 
Theory seminar led to my revisiting this material.}
Alexander Wilce \\ 
Department of Mathematics, Susquehanna University
\end{center}


\begin{abstract} 
The past {\red fifteen} years have seen a remarkable resurgence of the old programme of finding more or less a priori axioms for the mathematical 
framework of quantum mechanics. The new impetus comes largely from quantum information theory; in contrast to work in the older tradition, which tended to concentrate on structural features of individual quantum systems, the newer work is marked by an emphasis 
on systems in interaction. Within this newer work, one can discern two distinct approaches: one is ``top-down", and attempts to 
capture in category-theoretic terms what is distinctive about quantum information processing. The other is ``bottom up", attempting 
to construct non-classical models and theories by hand, as it were, and then characterizing those features that mark out quantum-like 
behavior.  This paper blends these approaches. We present a constructive, bottom-up recipe for building probabilistic theories 
having strong symmetry properties, using as data any uniform enlargement of the symmetric group $S(E)$ of any finite set, to a larger group 
$G(E)$. Subject to some natural conditions, our construction leads to a monoidal category of fully symmetric test spaces, in which 
the monoidal product is ``non-signaling". 
\end{abstract}

\section{Introduction}\label{intro}

After a long hiatus, there has been a recent resurgence of interest in 
axiomatic reconstructions or characterizations of quantum mechanics in probabilistic, or more broadly, informatic, terms. 
The new impetus comes largely from quantum information theory, and is marked by an emphasis, not on isolated physical systems and their properties, but on systems in interaction. Accordingly, the current focus is on characterizing (mainly, finite-dimensional) QM within a more general framework of abstract physical or probabilistic theories equipped with some device or devices for defining composite systems. 

At present, one can discern two approaches to this. The first (e.g., \cite{AC04,Baez06,BS09,Sel08}) is ``top-down": one  begins with a category of abstract physical systems, with arrows representing physical processes. This is generally assumed to be at least a symmetric monoidal category (and more usually, compact or dagger compact closed). In other words, it is assumed that there is a single, preferred method for composing systems ``in parallel".  

The second approach (e.g., \cite{BBLW07, BBLW08, BW08, Bar05, D'Ar09, FR81, Har01}), more explicitly probabilistic, is ``bottom-up": one first defines rather concretely what one means by an individual probabilistic model, and then introduces devices for combining and manipulating these, subject to a 
``non-signaling" constraint.  In place of a single, canonical tensor product, this approach provides a spectrum of possible ``non-signaling" products, bounded by a minimal product, allowing no entanglement between states, but arbitrary entanglement between effects, and a maximal product, allowing arbitrarily entangled states but no entangled effects. While this is adequate for discussing certain information-processing protocols (e.g., teleportation \cite{BBLW08, BW08}), if we are aiming at an axiomatic reconstruction or characterization of the usual apparatus of quantum mechanics, we need a unique tensor product, and one, moreover, that affords entanglement both between states and between effects. On the other hand, as the existence and uniqueness of such a product is presumably part of what one wants to explain, simply {\em postulating} it is ultimately  unsatisfactory: one should much prefer to {\em construct} the tensor product in some natural way. 

This paper takes a step in this direction. A conspicuous feature of both quantum and classical systems that has not been stressed in either of the approaches described above, is symmetry. Both classical and quantum systems are {\em homogeneous} in a strong sense: all pure states are alike, all basic measurements are alike, and all outcomes of such a measurement are alike. There is a simple construction  \cite{Wil05} whereby abstract probabilistic models having this same high degree of symmetry can be generated from a suitable extension of the symmetric group of a finite set $E$, representing the outcome-set of a basic experiment, to a larger group.   Where this construction can be made uniformly (that is, functorially), it leads to a probabilistic theory having a natural compositional structure. Both classical and quantum theory can be recovered in this way. In general, the composites arising from this construction need not satisfy all of the desiderata for a composite system in the sense of \cite{BBLW08, BW08, Bar05}. (In particular, there is a tension between requiring them to support arbitrary product states and arbitrary product measurements.) However, subject to 
a few simple and reasonably natural conditions, we are led to a symmetric monoidal category in which composite systems admit product measurements, and in which bipartite states are non-signaling.


\section{Probabilistic Models and Theories} 

There is a more or less standard mathematical framework for generalized probability theory, first sketched by Mackey 
\cite{Mackey} and later 
elaborated, modified, and in some instances, rediscovered, by many authors, e.g., \cite{D'Ar09, Davies-Lewis, FR72, Har01} 
The range of stylistic variation among these various formulations is just wide enough to make it prudent to spell out in a little detail the particular variant (one might say, dialect) 
 in which I'll proceed. In the interest of brevity, I consider here only the discrete, finite-dimensional version of this framework. 

In the language of \cite{BBLW08, BW08}, a finite-dimensional {\em abstract state space} is a pair $(A,u)$ where $A$ is a finite-dimensional ordered real vector space with positive cone $A_+$, and $u \in A^{\ast}$ is a distinguished {\em order unit}, i.e., a functional on $A$ that is {\em strictly} positive on $A$. The set $\Omega_{A} := \{ \alpha \in A | u(\alpha) = 1\}$ is the {\em normalized} state space. An {\em effect} on $A$ is  a positive functional $a \in A^{\ast}$ with $0 \leq a \leq u$ pointwise on $\Omega$; we regard $a(\alpha)$ as the probability of $a$ occurring when the state is $\alpha$. 
A {\em discrete observable} on $A$ is a set $E \subseteq V^{\ast}$ of effects with $\sum_{a \in E} a = u$. If $A$ is the self-adjoint part of a finite-dimensional complex $C^{\ast}$-algebra, i.e., a $\ast$-subalgebra of the algebra $M_{d}$ of $d \times d$ complex matrices, ordered as usual, and with $u(\alpha) = \tr(\alpha)/d$, then we may call $A$ a (finite-dimensional) {\em quantum} state space. 

For purposes of {\em constructing} such abstract models, it is often helpful (and clarifying) to introduce the following more operational apparatus, developed originally by D. J. Foulis and C. H. Randall in the service of quantum logic (see, e.g., \cite{FR81}).  

{\bf Definition 1:} 
A {\em test space} is a collection ${\mathfrak A}$ of
non-empty sets, called {\em tests}, understood as the outcome-sets
of various ``measurements". The set $X = \bigcup{\mathfrak A}$ of all outcomes of all tests is the {\em outcome space} for ${\mathfrak A}$. A {\em probability weight} on $\mathfrak A$ is a mapping $\alpha :
X \rightarrow [0,1]$ with $\sum_{x \in E} \alpha(x) = 1$ for all
$E \in {\mathfrak A}$. 

The convex set of all probability weights on $\mathfrak A$ is denoted $\Omega({\mathfrak A})$ for the convex set of all probability weights on ${\mathfrak A}$.  A {\em probabilistic model} is a pair $({\mathfrak A}, \Gamma)$, where $\mathfrak A$ is a test space and $\Gamma \subseteq \Omega({\mathfrak A})$ is a closed, compact, outcome-separating convex set of probability weights on ${\mathfrak A}$. As a default, we can always take $\Gamma = \Omega({\mathfrak A})$. This is the approach taken in most of the rest of this paper. When I refer to a test space as a model, this is what I have in mind. 

Given a probabilistic model $({\mathfrak A}, \Gamma)$, let $V = V({\mathfrak A},\Gamma)$ be the linear span of
$\Gamma$ in ${\mathbb R}^{X}$, ordered by the cone generated
by $\Gamma$. Letting $u \in V^{\ast}$ be the order unit corresponding to $\Omega$ (that is, 
the unique functional with $u(\alpha) = 1$ for all $\alpha \in \Gamma)$, the pair 
$(V,u)$ is then an abstract state space in the sense of [5,6]. Note that every outcome $x \in X$ induces a positive linear functional $f_x \in V^{\ast}$, given by $f_{x}(\omega) = \alpha(x)$ for all $\alpha \in \Gamma$. We have  $\sum_{x \in E} f_x = u$ for all $E \in {\mathfrak A}$, so $x \mapsto f_x$ is a discrete observable on $V$, in the sense of \cite{BBLW07}. (Thus, one can for many purposes regard a probabilistic model as an abstract state space equipped with a distinguished family of observables.)

From this point forward, I make the standing assumptions that (i) every test space $\mathfrak A$ is {\em locally finite}, that is, every test $E \in {\mathfrak A}$ is a finite set, and (ii) for every model $({\mathfrak A}, \Gamma)$, the space $V({\mathfrak A}, \Gamma)$ is finite-dimensional. 

{\bf Examples: classical and quantum models} (i) Let ${\mathfrak A} = \{E\}$
where $E$ is a finite set: then $\Omega({\mathfrak A})$ is the simplex $\Delta(E)$ of 
probability weights on $E$. If $\Hilb$ is a real, complex or quaternionic Hilbert space, the associated 
{\em quantum test space} is the set 
${\mathfrak F}(\Hilb)$ of orthonormal bases of $\Hilb$. Gleason's Theorem 
identifies $\Omega({\mathfrak F}(\Hilb))$ as the space $\Omega_{\Hilb}$ of 
density operator on $\Hilb$. 

{\bf Examples: Grids and Graphs} Here are two further examples that will figure importantly in the sequel. Fixing a set $E$, let $\Gr(E)$, the {\em grid} 
test space on $E$, be the set of rows and columns of $E \times E$, i.e., 
\[\Gr(E) = \{ \{x\} \times E | x \in E\} \cup \{ E \times \{y\} | y \in E\}.\]
Notice that a state on $\Gr(E)$ is essentially a $|E|$-by-$|E|$ doubly stochastic matrix. 

A related test space is the 
space 
\[\Gr(E)^{\ast} := \{ \Gamma_{f} : f \in S(E)\}\] of graphs $\Gamma_{f}$ of bijections $f : E \rightarrow E$. Equivalently, $\Gr(E)^{\ast}$ is the set of transversals of $\Gr(E)$, i.e., subsets of $E \times E$ meeting each row and each column exactly once (or, if we prefer, the space of supports of permutation matrices). Note that every test $\Gamma_{f} \in \Gr(E)^{\ast}$ induces a dispersion-free (that is, $\{0,1\}$-valued) state on $\Gr(E)$, and that every state on $\Gr(E)$ is a convex combination of these. Similarly, 
each row and each column of $\Gr(E)$ induces a dispersion-free on $\Gr(E)^{\ast}$. One can show that every state on $\Gr(E)^{\ast}$ is a convex combination of such row and column states.

\subsection{Products of Test Spaces} If $\mathfrak A$ and $\mathfrak B$ are test spaces, let ${\mathfrak A} \times
{\mathfrak B} = \{ E \times F | E \in {\mathfrak A}, F \in {\mathfrak B}\}$ be
the space of {\em product tests}. A state $\omega$ on ${\mathfrak A}
\times {\mathfrak B}$ is {\em non-signaling} if its {\bf marginal
states}
\[\omega_1 (x) := \sum_{y \in F} \omega(x,y)  \ \ \text{and} \ \ \omega_{2}(y) :=
\sum_{x \in E} \omega(x,y)\] are independent of $E \in {\mathfrak A}$
and $F \in {\mathfrak B}$, respectively. (The idea is that agents associated with $\mathfrak A$ and $\mathfrak B$ cannot then 
send information to one another by choosing to make one rather than another measurement.)

If $\alpha \in \Omega({\mathfrak A})$ and $\beta \in
\Omega({\mathfrak B})$, the {\em product state} 
\[(\alpha \otimes \beta)(x,y) := \alpha(x)\beta(y)\] 
is obviously non-signaling, as is any mixture of product 
states. In general, however, there will exist {\em entangled} non-signaling states that are {\em
not} mixtures of product states \cite{BBLW07, KRF87}. 

{\bf Definition 2:} A (non-signaling) {\em composite} of two test
spaces $\mathfrak A$ and ${\mathfrak B}$ is a test space ${\mathfrak C}$ plus an
embedding
\[ {\mathfrak A} \times {\mathfrak B} \rightarrow {\mathfrak C}\] such that
the restrictions of states on ${\mathfrak C}$ to ${\mathfrak A} \times {\mathfrak B}$ are non-signaling. 
If, in addition, every product state belongs to $\Omega({\mathfrak C})$, I'll call $\C$ a {\em full composite}.

Note that, by allowing ${\mathfrak C}$ to be larger than ${\mathfrak A} \times {\mathfrak B}$, we allow for the possibility of ``entangled" measurements, as well as entangled states.  By way of illustration, if $\Hilb_1$ and $\Hilb_2$ are complex Hilbert spaces, the test space ${\mathfrak F}(\Hilb_1 \otimes \Hilb_2)$ is a product of the test spaces ${\mathfrak F}(\Hilb_1)$ and ${\mathfrak F}(\Hilb_{2})$, under the embedding $(x,y) \mapsto x \otimes y$. In particular, all states on 
${\mathfrak F}(\Hilb_{1} \otimes \Hilb_{2})$. 


{\bf Example: The Foulis-Randall product} 
A minimal composite of test spaces, introduced by Foulis and Randall \cite{FR81}, is defined as follows. Given a test $E \in {\mathfrak A}$ and an $E$-indexed family of tests $F_x \in {\mathfrak B}$, the set $\bigcup_{x \in E} \{x\} \times F_x$ represents the outcome-set of a 
two-stage test, in which one first performs the test $E$ and then, upon securing $x \in E$, performs the test $F_x$. Let 
$\stackrel{\longrightarrow}{{\mathfrak A}{\mathfrak B}}$ denote the collection of all such two-stage tests, noting that 
${\mathfrak A} \times {\mathfrak B} \subseteq \stackrel{\longrightarrow}{{\mathfrak A}{\mathfrak B}}$, and also that these two test spaces have the same outcome-space, namely, $X({\mathfrak A}) \times X({\mathfrak B})$. Now let $\stackrel{\longleftarrow}{{\mathfrak A}{\mathfrak B}}$ denote the set of 
two-stage tests of the form $\bigcup_{y \in F} E_{y} \times \{y\}$ with $F \in {\mathfrak B}$ and $E_{y} \in {\mathfrak A}$ for every $y \in F$.
The {\em Foulis-Randall product} is 
\[{\mathfrak A}{\mathfrak B} \ := \ \stackrel{\longrightarrow}{{\mathfrak A}{\mathfrak B}} \cup \stackrel{\longleftarrow}{{\mathfrak A}{\mathfrak B}}.\] 
One can show that the state space $\Omega({\mathfrak A}{\mathfrak B})$ is exactly the set of non-signaling states on ${\mathfrak A} \times {\mathfrak B}$.  This product affords us no ``entangled outcomes", as outcomes of ${\mathfrak A}{\mathfrak B}$ are simply 
ordered pairs $(x,y)$ of outcomes $x \in X({\mathfrak A})$ and $y \in X({\mathfrak B})$. On the other hand, the easiest way to 
show that states on a test space ${\mathfrak C} \supseteq {\mathfrak A} \times {\mathfrak B}$ are non-signaling is to show that 
${\mathfrak C}$ contains all two-stage tests, i.e., that ${\mathfrak A}{\mathfrak B} \subseteq {\mathfrak C}$. I make use of this observation in the proof of {\red Theorem 1 in section 6}.

{\em Remark:} It is tempting to require, as a matter of definition, that states on a tensor product ${\mathfrak C}$ of test spaces $\mathfrak A$ and $\mathfrak B$ be {\em determined} by their restrictions to ${\mathfrak A} \times {\mathfrak B}$. 
When this condition --- called {\em local tomography} in \cite{BBLW07, BBLW08} --- is satisfied, 
When this condition is satisfied, conditions (i) and (ii) above guarantee that $\Omega({\mathfrak C})$ will be a tensor product, in the sense of \cite{BBLW07, BBLW08}, of the state spaces of $\mathfrak A$ and $\mathfrak B$, and, in particular, that $V({\mathfrak C})$ will be linearly isomorphic to $V({\mathfrak A}) \otimes V({\mathfrak B})$. However, 
this assumption is quite strong, being violated in real and quaternionic QM. 
For purposes of this paper, I prefer to keep to the more permissive definition above.

\subsection{Maps between test spaces} One can organize test spaces into a category in several different ways (for a more complete discussion, see \cite{Wilce09a}). An {\em event} of a test space ${\mathfrak A}$ is a subset of a test. That is, $A \subseteq X := \bigcup {\mathfrak A}$ is an event iff there exits some $E \in {\mathfrak A}$ with $A \subseteq E$. We write ${\cal E}({\mathfrak A})$ for the set of all events of $\mathfrak A$. Note that the empty set is an event, as is each test. (Indeed, if 
$\mathfrak A$ is irredundant, the tests are exactly the maximal events.) Naturally, we define the probability of an event $A$ in state $\alpha \in \Omega({\mathfrak A})$ by $\alpha(A) = \sum_{x \in A} \omega(x)$.  

Events $A, B \in {\cal E}({\mathfrak A})$ are {\em orthogonal}, written $A \perp B$, if they are disjoint and their union is an event. $A$ and $B$ are {\em complementary} iff they partition a test, i.e., $A \perp B$ and $A \cup B \in {\mathfrak A}$.  If $A$ and $B$ are both complementary to some event $C$, we say that $A$ and $B$ are {\em perspective}, with 
{\em axis} $C$, writing $A \sim B$ or $A \sim_{C} B$. Note that perspective events have the same probability in every state. Note, too, that any two tests are perspective, with axis the empty event.

{\bf Definition 3:} A {\em test space morphism} from a test space $\mathfrak A$ to a test space $\mathfrak B$ 
is a function $\phi : X({\mathfrak A}) \rightarrow X({\mathfrak B})$ taking events to events, and such that 
for all $A, B \in {\cal C}({\mathfrak A})$, 
\begin{itemize} 
\item[(i)] $A \perp B$ implies $\phi(A) \perp \phi(B)$, and 
\item[(ii)]$A \sim B$ implies $\phi(A) \sim \phi(B)$. 
\end{itemize}
Notice that, by (i), if $\beta$ is a state on ${\mathfrak B}$, then $\phi^{\ast}(\beta) := \beta \circ \phi$ is a 
state on ${\mathfrak A}$. 

It is straightforward that the composition of two morphisms (defined in the obvious way) is again a morphism, so we may speak of the {\em category} of test spaces and morphisms. Denote this category by $\Tesp$.
\footnote{A more general 
notion of test-space morphism allows for set-valued mappings; in \cite{Wilce09a}, the notation $\Tesp$ denotes 
the category of test spaces and these more general morphisms.}
 

\subsection{Connections with Quantum Logic} In the quantum-logical approach to generalized probability theory, one begins with an orthocomplemented poset --- usually, but not always, an orthmodular lattice or poset --- of ``propositions", treating states as probability measures on this structure. 
Test spaces provide (indeed, were invented in order to provide) a natural semantics for this approach \cite{FR81}. Perspectivity is obviously a symmetric and reflexive, but in general not a transitive, relation on events. On the other hand, in a quantum test space ${\mathfrak F}(\H)$, events (that is, orthonormal subsets of $\H$) are complementary iff they span orthogonal subspaces; hence, events are perspective iff they span the same subspace. In this case, then, perspectivity is an equivalence relation, and the quotient set ${\cal E}/\sim$ can be identified with the lattice $L(\H)$ of projection operators on $\H$. 

{\bf Definition 4:} 
A test space ${\mathfrak A}$ is {\em algebraic} iff perspective events in ${\cal E}({\mathfrak A})$ have exactly the same set 
of complementary events --- that is, if $A, B$ and $C$ are events with $A \sim B$ and $B \co C$, then $A \co C$. 

It follows that if $\mathfrak A$ is algebraic, $\sim$ is an equivalence relation on ${\cal E}$. We denote the equivalence class of $A \in {\cal E}({\mathfrak A})$ under perspectivity by $p(A)$; this is called the {\em proposition} associated with $A$. One can show that the quotient set ${\cal E}/\sim$ hosts a well-defined, associative partial binary operation defined by 
\[p(A) \oplus p(B) \ = \ p(A \cup B)\]
where $A$ and $B$ are complementary events. Equipped with this partial sum, ${\cal E}({\mathfrak A})/\sim$ is an 
{\em orthoalgebra} (see \cite{Wilce09a}), called the {\em logic} of ${\mathfrak A}$, and denoted $\Pi({\mathfrak A})$. This carries a natural partial order, given by $p(A) \leq p(B)$ iff $\exists C$ with $p(B) = p(A) \oplus p(C)$; this order is orthocomplemented by $p(A)' := p(C)$ where $C$ is any event complementary to $C$. Every orthoalgebra can be represented (canonically, though not uniquely) as the logic of a suitable test space. A morphism $\phi : {\mathfrak A} \rightarrow {\mathfrak B}$ between algebraic test spaces induces, in an obvious way (and in an obvious sense) an orthoalgebra homomorphism $\Pi(\phi) : \Pi({\mathfrak A}) \rightarrow \Pi({\mathfrak B})$, one can regard $\Pi$ as a functor from the category of algebraic test spaces and test-space morphisms to the category of orthoalgebras and orthoalgebra homomorphisms. 
 
Subject to various more-or-less reasonable (or at any rate, intelligible) constraints on the combinatorial structure of $\mathfrak A$, one can show that $\Pi({\mathfrak A})$ is variously an orthomodular poset, an orthomodular lattice, or a complete 
orthomodular lattice. Unfortunately, it seems to be  difficult to motivate algebraicity on operational grounds. Therefore, it is of interest to find other, more transparent conditions that imply algebraicity. One such condition is discussed in Section 6 below.

\section{Models with Symmetry}\label{models}

Let $G$ be a group. A $G$-{\em test
space} is a test space ${\mathfrak A}$ such that $X = \bigcup {\mathfrak A}$
carries a $G$ action, with $gE \in {\mathfrak A}$ for all $(g,E) \in G
\times {\mathfrak A}$ (so $G$ acts by {\em symmetries} of ${\mathfrak A})$.

{\bf Definition 5:} A $G$-test space ${\mathfrak A}$ is {\em fully $G$-symmetric} 
iff (i)
all tests have the same cardinality, and $(ii)$ any bijection $f : E \rightarrow F$ between
tests $E, F \in {\mathfrak A}$ is implemented by an element of $G$, in the sense that $f(x) = gx$ for all $x \in E$. 
Where this group element $g$ is uniquely determined, we say that ${\mathfrak A}$ is {\em strongly $G$-symmetric}. 

{\bf Examples:} Trivially, a classical test space is strongly 
symmetric under $S(E)$. The test space of frames of a real, complex or quaternionic Hilbert space $\Hilb$ is strongly, 
symmetric under the unitary group $U(\Hilb)$ of $\Hilb$
\footnote{If $\Hilb$ is an inner product space over {\em any} of the three classical division rings, 
I write $U(\Hilb)$ for the group of isometries of $\Hilb$, i.e., linear mappings $u : \Hilb \rightarrow \Hilb$ 
with $\langle u x, u y \rangle = \langle x, y \rangle$ for all $x, y \in \Hilb$. }
The space of {\em projective} frames, i.e, 
maximal families of rank-one projections on $\Hilb$, is fully but not strongly $U(\Hilb)$-symmetric, as a bijection 
between projective frames determines a unitary only up to a choice of a phase for each $x \in E$. 
Both $\frak{Gr}(E)$ and $\Gr(E)^{\ast}$ are fully symmetric: the former under the subgroup
of $S(E \times E)$ generated by row shifts, column shifts and
transpose; the latter under row and column shifts alone (i.e.,
$S(E) \times S(E)$ acting by $(\sigma, \tau)(x,y) = (\sigma x, \tau
y)$). 

As a rule, highly symmetric mathematical objects can be reconstructed from knowledge of their symmetries. Fully symmetric test spaces are no exception:

\subsection{Basic Construction} 
Let $H$ be a group, and let $E$ be an $H$-set, that is, a set upon which $H$ acts. One might think of $E$ as representing a prototypical experiment, singled out for reference, and $H$ as a preferred group of symmetries of $E$. Say that $H$ acts {\em fully} on $E$ iff the action $H \rightarrow S(E)$ is surjective, so that every permutation of $E$ is implemented by some $h \in H$.  {\red Then,} in particular, $E$ is a transitive $H$-set, so $E \simeq H/H_{x_o}$, where $H_{x_o}$ is the stabilizer of any chosen base-point $x_o \in E$. Now, fixing $x_o$, let $G$ be a group extending $H$, in the sense that $H \leq G$, and let $K \leq G$ with 
\begin{equation} K \cap H = H_{x_o}. \end{equation}
Let $X := X(G,H,K) = G/K$, understood as a 
$G$-set; let $\phi : E \rightarrow X$ be given by $\phi(x) = hK$ where $x = h x_o \in E$. Condition (\theequation) guarantees that $\phi$ is a well-defined, $H$-equivariant injection. Henceforth, we identify $E$ with its image under $\phi$, understanding 
$E$ as an $H$-invariant subset of $X$. Finally, let ${\mathfrak A} = {\mathfrak A}(G,H,K)$ be the orbit of the set $E \subseteq X$ under the action of $G$, i.e., 
\[{\mathfrak A}(G,H,K) \ = \ \{~ gE ~|~ g \in G ~\}.\] Note that $\bigcup {\mathfrak A} = X$. To see that ${\mathfrak A}$ is fully $G$-symmetric, 
let $f : gE \rightarrow g'E$ be any bijection between two tests in ${\mathfrak A}$. Then $(g')^{-1} \circ f \circ g : E \rightarrow E$ defines a permutation of $E$; hence, there is some $h \in H$ with $(g')^{-1} f (gx) = hx$ for every $x \in E$, whence, $f(y) = g' h g^{-1} y$ for every $y \in gE$. 

{\em Remarks:} \\
(1) Given $G,H$ and $K$ as above, we can {\em define} $E = H/(H \cap K)$. Thus, in principle the construction depends only on purely group-theoretic data: a group $G$ and a pair of subgroups $H, K \leq G$. 

(2) Note that, in the foregoing construction, we made no real use of the fact that $H$ acts {\em fully} on $E$: any transitive action would have done as well. We will make no use here of this extra generality, but it's worth bearing in mind its availability. 

If $\mathfrak A$ is a fully symmetric $G$-test space, let $H = G_{E}$, the stabilizer of a test $E \in \mathfrak A$, 
and let $K = G_{x}$ where $x \in E$. Then ${\mathfrak A}$ is canonically isomorphic to ${\mathfrak A}(G,H,K)$. As an 
illustration, let ${\mathfrak A} = {\mathfrak F}(\H)$ for a (say, complex) Hilbert space $\H$, and let $G$ be the unitary 
group $U(\H)$. Then $H = G_{E}$ is the set of unitaries permuting the frame $E$, while $K = G_x$ is the set of unitaries 
fixing the unit vector $x \in E$. Then $X = G/G_{x}$ can be identified with the unit sphere of $\H$, and 
${\mathfrak A}(G,H,K)$ is the orbit of the frame $E$ --- that is, the set ${\mathfrak F}(\H) = {\mathfrak A}$ of all orthonormal 
frames.



We call a {\red probabilistic} model $({\mathfrak A}, \Gamma)$ fully symmetric (under $G$) iff $\mathfrak A$ is fully symmetric, 
$\Gamma$ is invariant under $G$'s natural action on ${\mathbb R}^{X}$, and $G$ acts transitively on 
the extreme points of $\Gamma$.  Note that if $\mathfrak A$ is a fully symmetric $G$-test space and 
$\alpha_o$ is a chosen state in $\Omega({\mathfrak A})$, we obtain a fully symmetric model by taking $\Gamma$ 
to be the convex hull of the orbit of $\alpha_o$ under $G$. In all four cases considered above, the full state space is invariant, and extreme states are permuted transitively, so these models are already fully symmetric. 

\subsection{Linear Representations} That it be fully symmetric does not, by itself, guarantee that a model will be very interesting. 
In particular, a fully symmetric test space need not have very many states. As an example, consider the test space 
$\{\{a,b\}, \{b,c\}, \{c,a\}\}$: this is obviously fully symmetric under the group $S_3$, but has  only 
one state, namely, $\alpha(a) = \alpha(b) = \alpha(c) = 1/2$. 

On the other hand, if a fully symmetric test space {\em is} endowed with 
a rich state space, good things follow.  Let ${\mathfrak A}$ be a fully $G$-symmetric test space,  $G$  a compact 
group. Let $V = V({\mathfrak A})$ be the ordered vector space spanned by $\Omega({\mathfrak A})$, as discussed in section 2. 
Fixing an outcome $x_o \in X$, we have a surjection $G \mapsto X
= \bigcup{\mathfrak A}$ given by $g \mapsto gx_o$, and hence, dualizing, an
embedding $V \rightarrow C(G)$ 
of the state space of $\mathfrak A$ in the algebra of continuous real-valued functions on $G$, 
given by $\omega \mapsto \hat{\omega}(g) := \omega(gx_o)$.  One easily verifies that the cone $V_{+}$, thus embedded, 
is closed under convolution; hence, we may regard $V$ as a sub-algebra of ${\mathbb C}[G]$. 
This gives us 
an invariant inner product on $V$, which is {\em positive} in the sense 
that $\langle \alpha, \beta \rangle \geq 0$ for all $\alpha, \beta \in V_{+}$. Using this, one can show \cite{QW08} that if a fully-$G$-symmetric test space 
${\mathfrak A}$ has a separating, finite-dimensional state space, then $V^{\ast}$ can be endowed with a $G$-invariant inner product, positive on the positive cone of $V$, and ${\mathfrak A}$ can be represented as an invariant family of orthonormal subsets of $V^{\ast}$.


\section{Fully Symmetric Theories}\label{recipe} 

If our goal is to construct and study, not individual probabilistic models, but probabilistic {\em theories} -- classes, or better, 
categories, of such models -- then we might consider uniformizing the construction $H,K \leq G \Rightarrow {\mathfrak A}(G,H,K)$ described above. In this section, I consider one way of doing this. In the interest of simplicity, I consider only the case in which $H$ is the symmetric group of a typical test. 


In order to make the standard construction of Section 3.1 in a uniform way, we should like to associate to every finite set $A$ a group $G(A)$ and a fixed embedding $j_{A} : S(A) \rightarrow G(A)$, in such a way that 
\begin{equation} 
A \subseteq B \ \Rightarrow \ G(A) \leq G(B) \ \text{and} \ G(A) \cap S(B) = S(A).\end{equation}
This suggests treating $S$ and $G$ as {\em functors} from an appropriate category of sets into the category 
of groups, and $j : A \mapsto j_{A}$ as a natural transformation from $S$ to $G$. Now, the assignment 
$A \mapsto S(A)$ of a set to its symmetric group is {\em not} the object part of any sensible functor from the category $\Set$ of finite sets and arbitrary mappings to the category $\Grp$ of groups and homomorphisms, but it {\em is} functorial in the category $\Sinj$ of finite sets and {\em injective} mappings: if $f : A \rightarrow B$ is an injection, we have a natural homomorphism $S(f) : S(A) \rightarrow S(B)$ given by 
\[S(f)(\sigma)(b) = \left \{ \begin{array}{cl} f(\sigma(a)) & b = f(a) \\ b & b \not \in \ran(f) \end{array} \right . \]
Note that where $i : A \subseteq B$ is an inclusion, we have $S(i)(\sigma)(a) = \sigma a$ for all $a \in A$ and 
$S(i)(\sigma) b = b$ for every $b \in B \setminus A$, i.e., $S(i)$ is the standard embedding of $S(A)$ as a subgroup of $S(B)$. I'll routinely identify $S(A)$ with its image under this embedding, writing $S(A) \leq S(B)$. 

Suppose now that $j : S \rightarrow G$ is a natural transformation from $S$ to a functor 
$G : \Sinj \rightarrow \Grp$, so that we have for every object $A \in \Sinj$, a homomorphism $j_{A} : S(A) \rightarrow G(A)$, 
such that for every injection $f : A \rightarrow B$, the square 
\begin{equation} {\xymatrix@=12pt{ S(A) \ar@{->}_{S(f)}[dd] \ar@{->}^{j_{A}}[rr]&  & G(A) \ar@{->}^{G(f)}[dd]  \\
& & & \\\
S(B) \ar@{->}_{j_{B}}[rr] & & G(B) }}\end{equation}
commutes. 
In order to guarantee that condition (2) is satisfied, I make the following 

{\bf Definition 6:} An {\em extension} of the functor $S : \Sinj \rightarrow \Grp$ is a pair $(G,j)$ where $G$ is a functor from 
$\Sinj$ to $\Grp$, $j : S \rightarrow G$ is a natural transformation from $S$ to $G$, and, for 
every injective mapping $f : A \rightarrow B$, 
{\begin{mlist} 
\item[(i)]  $G(f) : G(A) \rightarrow G(B)$ is injective, and   
\item[(ii)]  the square (\theequation) is a pull-back.
\end{mlist}}

Where  $A \subseteq B$, the inclusion mapping $i : A \rightarrow B$ provides a canonical embedding $G(i) : G(A) \rightarrow G(B)$. Identifying $G(A)$ with its image under $G(i)$, I'll regard $G(A)$ as a subgroup of $G(B)$. I'll also 
identify $S(A)$ with its image under $j_{A}$, writing $S(A) \leq G(A)$. With these conventions, we have 

{\bf Lemma 1:} {\em Let $A \subseteq B$. Then $G(A) \cap S(B) = S(A)$.} 

{\em Proof:} Let $i : A \subseteq B$ be the inclusion mapping. The left hand side above is more exactly 
$G(A) \cap S(B) = G(i)(G(A)) \cap j_{B}(S(B))$; the right hand side is $G(i) j_{A}(S(A))$. Since 
$G(i) \circ j_{A} = j_{B} \circ S(i)$, the right hand side is contained in the left. Let's verify this explicitly. 
If $\sigma \in S(A)$, we have 
\[S(A) = G(i) j_{A}(S(A)) = j_{B} S(i)(S(A)) \subseteq j_{B}(S(B)).\]
We also have 
\[G(i) j_{A}(\sigma) \in G(i) j_{A}(S(A)) = G(i) G(A) = G(A) \leq G(B).\]
So $S(A) \subseteq S(B) \cap G(A)$. Conversely, let $g \in S(B) \cap G(A)$. Then 
$g = j_{B}(\sigma)$ for some $\sigma \in S(B)$. Now $g \in G(A)$, so $g = G(i)(g')$ 
for $g' \in G(A)$. Since the square is a pullback, there exists $\sigma' \in S(A)$ 
with $\sigma = G(i)(\sigma')$ -- i.e., $\sigma' = \sigma$ -- and $j_{A}(\sigma) = g'$. 
So, by commutativity of the square, $g = G(i) j_{A} (\sigma) \in S(A) \leq G(B)$. $\Box$

Now for $A \not = \emptyset$, fix a base point $a \in A$, and set $K(A,a) = G(A \setminus a)$. The Basic Construction of section 3 yields a fully $G(A)$-symmetric 
test space 
\[{\mathfrak G}(A,a) \ := \ {\mathfrak A}(G(A),S(A),K(A,a))\]
with outcome-space 
\[X(A,a) \ := \  G(A)/K(A,a),\] 
and a canonical, $S(A)$-equivariant embedding $A \rightarrow X(A,a)$, with  
${\mathfrak G}(A,a)$ the orbit of $A$ in $X(A,a)$, so that each test has the form $gA$ for some $g \in G(A)$.

Note that we already have a candidate for a canonical ``tensor product" of $\G(A,a)$ and $\G(B,b)$, namely, 
\[{\mathfrak G}(A,a) \otimes {\mathfrak G}(B,b) := {\mathfrak G}(A \times B, (a,b)).\]
However, as we'll now see, without some further restrictions on the extension $(G,j)$, this may exhibit some rather pathological (or, depending on one's taste, rather interesting) behavior.

\subsection{Three Examples} 

{\red Where the choice of base-point is unimportant, we can abbreviate $X(A,a)$ and $\G(A,a)$ as 
$X(A)$, $\G(A)$, respectively --- bearing in mind that such a choice has been made, tacitly. 
I follow this convention in the following examples. (In Section 5, I'll impose a further condition 
on $(G,j)$ that will establish canonical isomorphisms $X(A,a) \simeq X(A,a')$ for all $a, a' \in A$, 
making this practice respectable.)}

{\bf Example: Unitary Extensions} We can regard the passage from $S$ to $G$, and the associated passage from $\Sinj$ to $G-\Tesp$, as a kind of abstract {\em quantization rule}. Indeed, there is a natural functor $U : \Sinj \rightarrow \Grp$ assigning to each (finite) set $A$ the unitary group $U(A)$ of the finite-dimensional Hilbert space {\red $\H(A) := ({\mathbb C}^{A})^{\ast}$. Identifying $A$ with its canonical image in $\H(A)$, each injection $f : A \rightarrow B$ 
extends uniquely to a unitary embedding $u_f : \H(A) \rightarrow \H(B)$. }
This, in turn, gives us a homomorphism $U(f) : U(A) \rightarrow U(B)$, defined by $U(f)(g) = u_f g u_{f}^{\ast} \oplus \1_{B-\ran{f}}$, where 
$\1_{B -\ran{f}}$ is the identity operator on $\H(B \setminus \ran(f))$.  
It is easy to check that $U$ extends $S$ in the desired way (noting that a permutation matrix is a special kind of unitary). Applying the recipe above, {\red the resulting $U(A)$-test space $(X(A),{\mathfrak U}(A))$ is, up to obvious isomorphisms, the quantum test space $(X(\H(A)), {\mathfrak F}(\H(A))$.  Notice that 
$\mathfrak U(E) \otimes {\mathfrak U}(F) := {\mathfrak U}(E \times F)$ is canonically isomorphic to the test space of frames of $\H(E) \otimes \H(F)$.  }



We now consider the ``grid" and ``graph" test test spaces of Section 2 in this light. 
 
{\bf Example: The Grid Extension} Let $G(A)$ be the subgroup of $S(A \times A)$ generated by $S(A) \times S(A)$ together with the 
transposition mapping $\tau_A : (x,y) \mapsto (y,x)$; 
and let $j_{A}(\sigma) = (\sigma, \id_{A})$. For $f : A \rightarrow B$, let $G(f) : G(A) \rightarrow G(B)$ be the homomorphism 
determined by $G(f)(\sigma_1, \sigma_2) = (S(f)(\sigma_1), \sigma_2)$ and $G(\tau_A) = \tau_B$. 
One can work out that, for this extension, $X(A) = A \times A$ (up to choice of base-point), and ${\mathfrak G}(A) = \frak{Gr}(A)$, the grid test space considered above.  Thus, we have 
\[\frak{Gr}(A) \otimes \frak{Gr}(A) = \frak{Gr}(A \times B).\] Observe that $\frak{Gr}(A \times B)$ has arbitrary product states (essentially, because the cartesian product of two permutations is a permutation), but 
{\em lacks} arbitrary product {\em tests}: row-times-row and column-times-column tests are well-defined 
members of $\Gr(A \times B)$, but if $E$ is a row of ${\Gr}(A)$ and $F$, 	 a column, then the row-times-column 
set $E \times F$ is neither a row nor a column of $E \times F$ (it is, rather, a block sub-grid of the latter).  
Moreover, states on ${\mathfrak Gr}(A \times B)$ are typically {\em signaling} (essentially, because there is a 
correlation between which measurements on the second factor are available, depending upon which measurement is made on the first factor.) So this is {\em not} a composite in the sense of Definition 2.

{\bf Example: The Graph Extension} Let $G(A) = S(A) \times S(A)$, and embed $S(A)$ in $G(A)$ by
$j_{A}(\sigma)  = (\sigma, \sigma)$. If $f : A \rightarrow B$ is an injection, let 
$G(f) = S(f) \times S(f)$ Then ${\mathfrak G}(A) = \Gr(B)^{\ast}$, and
\[\Gr(A)^{\ast} \otimes \Gr(B)^{\ast} = \Gr(A \times B)^{\ast}.\] 
Let $\lambda : (A \times B)^2  \rightarrow A^{2} \times B^2$ be the map
$\lambda : ((x,y),(u,v)) \rightarrow ((x,u),(y,v))$: 
one can check that $\lambda(\Gamma_f \times \Gamma_g) = \Gamma_{f
\times g}$ for $f, g \in \Gr(E)^{\ast}$, so we have a natural
mapping $\lambda : \Gr(A)^{\ast} \times \Gr(B)^{\ast} \rightarrow
\Gr(A \times B)^{\ast}$.  
States on $\Gr(A \times B)^{\ast}$ pull back
along $\lambda$ to non-signaling states on $\Gr(A)^{\ast} \times
\Gr(B)^{\ast}$. So this is closer to being a product according to our
previous definition. However, there is still a problem: arbitrary products of {\em states} on 
$\Gr(A)^{\ast}$ {\red and $\Gr(B)^{\ast}$} need not be states on $\Gr(A \times B)^{\ast}$: the product of a row state and a {\red on $\Gr(A)^{\ast}$ and a column state on $\Gr(B)^{\ast}$}, for instance, will not be a convex combination of row or column states on $\Gr(A \times B)^{\ast}$, 
and hence, will not be a state on the latter. 

The moral seems to be that, for fully symmetric theories, there is a certain tension
between the demand for arbitrary product states, and the demand for
arbitrary product measurements.  Of course, if we want to view $\G(A \times B)$ as a composite of 
$\G(A)$ and $\G(B)$ in the sense of Definition 2, we need to find a natural mapping $X(A) \times X(B) \rightarrow X(A \times B)$. Later sections of this paper will investigate conditions on $(G,j)$ that will allow us to define such a 
mapping. As we'll see (Theorem 1), these conditions --- which force $\G(A \times B)$ to contain product 
tests --- will exclude the ``Grid" example.

{\blue \subsection{The category $G$-$\Tesp$}}

If $f : A \rightarrow B$ be an injection, $a \in A$, and  $b = f(a) \in B$, then 
$f|_{A \setminus a}$ takes $A \setminus a$ to $B \setminus b$; hence,  $G(f)$ takes  
 $K(A,a) = G(A \setminus a)$ into $K(A,b) = G(B \setminus b)$. 
It follows that $f : A \rightarrow B$ gives rise, to a well-defined map 
\[X_{a}(f) : X(A,a) \rightarrow X(B,f(a))\]
given by 
\[X_{a}(f)(ga) = G(f)(g)f(a).\] 
for all $g \in G(A)$. 
{\red One can also check that if $f : A \rightarrow B$ and $g : B \rightarrow C$ are injections with 
$f(a) = b$ and $g(b) = c$, then 
\[X_{a}(g \circ f) = X_{b}(g) \circ X_{a}(f).\]
In other words, we have here a functor $X$ from the category of pointed finite sets and injective (point-preserving) maps, 
to test spaces. }

It should be noted that, at this level of generality, $X_a(f)$ need not be a test-space morphism from ${\mathfrak G}(A,a)$ to ${\mathfrak G}(B,f(a))$ (though this will be the case if $(G,j)$ satisfies an additional condition, discussed below in Section 5). We can nonetheless define a category, 
which I'll call $G-\Tesp$, having as its objects test spaces of the form ${\mathfrak G}(A,a)$, and as its morphisms, composites 
of maps of the form $X_a(f)$ and symmetries $g \in G(A)$ --- so that, for instance, given injections $f_1 : A \rightarrow B$, $f_2 : B \rightarrow C$, base-points $a$, $b$ and $c$ with $c = f_2(b), b = f_1(a)$, and group elements $g \in G(A), h \in G(B)$ and $k \in G(C)$, $k \circ X_{b}(f_2) \circ h \circ X_{a}(f_1) \circ g : X(A,a) \rightarrow X(C,c)$ is a $G-\Tesp$ morphism. By the {\em theory associated with} an extension $(G,j)$, I'll mean this category.

{\bf Example:} Let $U$ be the unitary extension discussed above. Recall that 
$U(A)$ is the unitary group of $\H(A) = (\C^{A})^{\ast}$. As discussed earlier, 
every injection $f : A \rightarrow B$ gives rise to unitary embedding 
$u_f : \H(A) \rightarrow \H(B)$.  Suppose now that $u : \H(A) \rightarrow \H(B)$ is a unitary embedding. Let $B' = u(A) \subseteq X(B)$, and let $g \in U(B)$ be any unitary with $gB' = B$; then we have a map $g \circ u|_{A} : A \rightarrow B$, and hence, a unitary embedding $X(g u|_{A})$; since this agrees with $g \circ u$ on $A$, an orthonormal basis for $\H(A)$, these two unitary maps are the same; hence, $u = g^{-1}X(gu_{A})$.  Thus, the category $U-\Tesp$ is just the category of finite-dimensional complex Hilbert 
spaces (more exactly, but irrelevantly: such spaces with preferred orthonormal bases), and unitary embeddings.

{\red 
\subsection{Injectivity of $X_{a}(f)$}

If $A \subseteq B$ and $i_{A,B} : A \rightarrow B$ is the inclusion mapping, we have a canonical mapping 
$X_{a}(i_{A,B}) : X(A,a) \rightarrow X(B,a)$ for any $a \in A$. This takes $A$ injectively to $B$, but need not be injective on all of $X(A,a)$. Indeed, 

{\bf Lemma 2:} {\em The following are equivalent:
\begin{itemize} 
\item[(a)] $X_{a}(i_{A,B})$ is injective, for all $a \in A \subseteq B$
\item[(b)] $G(A) \cap G(B \setminus a) = G(A \setminus a)$ for all $a \in A \subseteq B$; 
\item[(c)] $G(A \cap B) = G(A) \cap B(B)$ for all $A, B$
\end{itemize} }

{\em Proof:} The definition of $X_a$, as applied to the inclusion mapping $i_{A,B}$, gives us 
$X_a(i_{A,B})(gK(A,a)) = gK(B,a)$. 
$X_a(i)$ is injective iff, for all $g \in G(A)$, $gK(B,a) = K(B,a)$ implies $gK(A,a) = K(A,a)$, 
i.e., iff $G(A) \cap G(B \setminus a) \subseteq G(A \setminus a)$. Since $G(A \setminus a) \subseteq 
G(A) \cap G(B \setminus a)$ in any case, we have that (a) implies (b). 
Clearly, (c) implies (b), 
so it remains to show that (b) implies (c). 

Suppose, then, that (b) holds for all choices of $a \in A \subseteq B$. I claim that 
$G(A) \cap G(B \setminus D) = G(A \setminus D)$ for all $D \subseteq A$. Let $D = \{a_o,a_1,..,a_n\} \subseteq A$, 
and let $D_k = \{a_o,...a_k\}$ for all $k = 0,...,n$. Since $G(B \setminus D) \subseteq G(B \setminus D_k)$ 
for all $k$, we have $G(B \setminus D) = \bigcap_{k=0}^{n} G(B \setminus D_k)$. Now apply (b) iteratively: 
\begin{eqnarray*}
G(A) \cap G(B \setminus D) & = & \bigcap_{k} G(A) \cap G(B \setminus a_o) \cap G(B \setminus \{a_o,a_1\}) \cap 
\cdots \cap G(B \setminus D)\\
& = & G(A \setminus a_o) \cap G(B \setminus \{a_o,a_1\}) \cap \cdots \cap G(B \setminus D)\\
& = & G(A \setminus \{a_o,a_1\}) \cap G(B \setminus \{a_o,a_1,a_2\}) \cap \cdots \cap G(B \setminus D)\\
& \vdots & \\
& = & G(A \setminus D) 
\end{eqnarray*}
Now let $D = A \setminus B$: 
\[\ \ G(A \cap B) = G(A \setminus (A \setminus B)) = G(A) \cap G(B \setminus (A \setminus B)) = G(A) \cap G(B). \ \Box \]

In all of the examples considered above, it is in fact true that 
$G(A \cap B) = G(A) \cap G(B)$. However, this need not hold in general. 
The following is due to Peter Selinger\footnote{Personal communication}. 

{\bf Example:} Let $F$ be the endofunctor of $\FinInj$ given by $F(A) = A \oplus \{0,1\}$ if $|A| > 0$, and $F(\emptyset) = \emptyset$, 
with $F(f) : F(A) \rightarrow F(B)$ the obvious injection if $f$ is an injection $A \rightarrow B$. 
Define $G(A) = S(F(A))$, with $j_{A} : S(A) \rightarrow G(A)$ the mapping $S(i_A)$, where $i_A : A \rightarrow F(A)$ is 
the inclusion mapping. 
The resulting natural transformation 
is cartesian (since $F$ is cartesian and $S$ preserves pullbacks), so we have an extension in the sense of Definition 6. 
However, if $A$ and $B$ are disjoint, we have 
$G(A \cap B) = S(\emptyset)$ while $G(A) \cap G(B) = S(\{0,1\})$. 



It will greatly simplify matters (in particular, the notation!) to assume --- and henceforth, I shall assume --- 
that $G$ satisfies the equivalent conditions (a)-(c) above.  Since $X_a(f)$ is bijective whenever $f$ is bijective, 
An immediate consequence is 

{\bf Corollary 1:} {\em Subject to conditions (a)-(c) of Lemma 2, $X_a(f) : X(A) \rightarrow X(B)$ is injective 
for every injection $f : A \rightarrow B$.}


 } 

\section{Regular Extensions} 

{\red Since we are identifying $\sigma \in S(A)$ with its image $j_{A}(\sigma) \in G(A)$, it will be helpful below to use the notation $\sigma|_{A}$ to refer to $\sigma$ in its original role as a bijection $A \rightarrow A$. 
Applying the functor $G$ gives us a homomorphism $G(\sigma|_A) : G(A,a) \rightarrow G(A,\sigma a)$, and also a 
mapping  $X(\sigma|_{A}) : X(A, a) \rightarrow X(A, \sigma a)$ --- injective, given our standing assumption that  $G$ 
satisfies the conditions of Lemma 2. We have $X(\sigma|_A)(x) = \sigma x$ for all $x \in A$; it is 
natural to ask that $X(\sigma|_A)$ coincide with $\sigma$'s action on $X(A,a)$, in the sense that 
$X(\sigma|_A)(g a) = \sigma g a = (\sigma g \sigma^{-1})(\sigma a)$ for all $g \in g(A)$. This will 
follow if $G(\sigma)$ is conjugation by $\sigma$, i.e., $G(\sigma)(g) = \sigma g \sigma^{-1}$. However, 
it will turn out to be very useful to impose a slightly stronger condition: }

{\bf Definition 7:} An extension $(G,j)$ is {\em regular} iff for all finite sets $A$, and for all $\sigma \in G(A)$ with $\sigma A = A$ -- that is, 
for all $\sigma$ in the stabilizer, $G(A)_{A}$, of $A$ in $G(A)$ -- we have  $G(\sigma{\red |_{A}})(g) = \sigma g \sigma^{-1}$ 
for all $g \in G$. 

It is easy to check 
that the unitary extension $(U,j)$ and the ``graph" extension $G(A) = S(A) \times S(A)$, $j_{A}(\sigma) = (\sigma, \sigma)$ are regular. The ``grid" extension, in which $G(A)$ is the subgroup of $S(A \times A)$ generated by $G(A) \times G(A)$ and transposition, with 
$j_{A}(\sigma) = (\sigma, \id_{A})$, is not regular (since if $\sigma \in S(A)$, 
$G(\sigma|_{A})(\tau_A) = \tau_A \not = (\sigma,\id_A) \tau_A (\sigma^{-1}, \id_A) = \tau_C (\sigma^{-1}, \sigma)$).

{\bf Lemma {\red 3}:} {\em Let $(G,j)$ be a regular extension. Then, for every finite set $A$ and 
base-point $a \in A$, $X_a(\sigma|_{A})x = \sigma x$ for every $\sigma \in G(A)_{A}$ and every $x \in X(A)$.}

{\em Proof:} For any $a \in A$, we have 
\[X_a(\sigma|_{A})(ga)  = G(\sigma_{A})(g) \sigma a = \sigma g \sigma^{-1} \sigma a = \sigma g a\] 
for all $g \in G$ --- hence, $X_a(\sigma|_{A})x = \sigma x$ for all $\sigma \in G(A)_{A}$ and all $x \in X(A)$. $\Box$ 

{\green The following will be useful later, in section 6. 

{\bf Lemma 4:} {\em If $A \subseteq B$, let $F(A)$ be the subgroup of $G(B)$ fixing $A$ pointwise. 
If $(G,j)$ is regular, we have 
\[F^{B}_{A} = G(B \setminus A).\]}

{\em Proof:} Note that $F^{B}_{b} = G(B \setminus b)$ by definition (where $b$ is the chosen basepoint used to 
construct $X(B)$). Now suppose $c = \sigma b$ for some $\sigma \in S(B)$. The functoriality of $G$ gives us 
\[G(\sigma_B)G(B \setminus b) = G(B \setminus c).\]
Since $G$ is regular, 
\[G(\sigma)G(B \setminus b) = \sigma G(B \setminus b) \sigma^{-1} = \sigma F^{B}_{b} \sigma^{-1} = 
F^{B}_{\sigma b} = F^{B}_{c}.\]
In other words, 
\[G(B)_{c} = G(B \setminus c)\]
for all $c \in B$. 
Now proceed inductively: 
\[F^{B}_{c_1, c_2} = (F^{B}_{c_1})_{c_2} = G(B \setminus c_1)_{c_2} = G((B \setminus c_1) \setminus c_2) = G(B \setminus \{c_1, c_2\})\]
and so on. $\Box$
}

Note the important special case where $B = A$: here $F^{A}_{A} = G(A \setminus A) = G(\emptyset)$. It follows 
that any $k \in G(\emptyset)$ fixes every point of $A$. Hence, in particular, $k \in G(A)_{A}$, whence, by the definition of regularity, we have 
\[k g k^{-1} = G(k|_{A})(g) = S(\id_{A})(g) = g.\]
It follows that $kg = gk$ for every $g \in G(A)$. Hence, for any $gb \in X(A)$, $kgb = gkb = gb$. In 
other words, $k$ fixes every element of $X(A)$. Conversely, 
if $k$ fixes every element of $X(A)$, it also fixes $A$ pointwise, whence, belongs to $F^{A}_{A} = G(\emptyset)$. 
It follows that if $g \in G(A)_{A}$, then letting $\sigma = g|_{A} \in S(A)$, we have $\sigma^{-1} g \in F^{A}_{A} = G(\emptyset)$, whence, $g = \sigma k$ for some $k \in G(\emptyset)$. 

It will be helpful to record these observations as

{\bf Lemma 5:} {\em Let $(G,j)$ be regular.  Then for every finite set $A$, 
\begin{itemize} 
\item[(a)] $k \in G(\emptyset) \leq G(A)$ iff $k$ fixes every element of $X(A)$; 
\item[(b)] $g \in G(A)_{A}$ iff $g = \sigma k$ for some $\sigma \in G(A)$ and $k \in G(\emptyset)$.
\end{itemize}}

Regularity also provides us with a {\em canonical} isomorphism $X(A,a) \simeq X(A,b)$ for all $a,b \in A$. 
Indeed, if $b = \sigma a$ for some $\sigma \in S(A)$, then we have 
\[K(A,b) := G(A \setminus b) = G(\sigma|_A)(G(A \setminus a)) = \sigma K(A,a) \sigma^{-1}.\]
This last is the stabilizer of $b$, so we have a canonical $G(A)$-equivariant 
mapping $\phi : X(A) \simeq G(A)/K(A,b) = X(A,b)$, given by
{\blue 
\[\phi(gK(A,a) = g \sigma^{-1} K(A,b),\]
where $\sigma$ is any element of $G(A)$ taking $a$ to $b$. (To see that this is independent of $\sigma$:  if $\sigma' a = \sigma a = b$, then $\sigma' = k\sigma$ where 
$k$ is in the stabilizer of $b$, i.e, $k \in K(A,b)$; thus, $g {\sigma'}^{-1} K(A,b) = g \sigma^{-1} k^{-1} K(A,b) = g \sigma K(A,b)$.)

It follows that, where $(G,j)$ is regular, for an injection $f : A \rightarrow B$, the mapping $X_a(f) : X(A,a) \rightarrow X(A,f(a))$ doesn't really depend on the choice of base-point.  
In other words, if $\phi : X(A,a) \simeq X(A,b)$ and $\psi : X(B,f(a)) \simeq X(B,f(b))$ are the canonical isomorphisms, we want to show that, for all $g \in g(A)$, 
\[\psi (X_{a}(f)(ga)) = X_{b}(f)(\phi(ga)).\]
Letting $\sigma a = b$ as before, let $\tau = S(f)(\sigma)$, so that 
we have $\tau f(a) = S(f)(\sigma)(f(a)) = f(\sigma a) = f(b)$. Then  
\begin{eqnarray*}
X_{b}(f)(\phi(ga)) & = & X_{b}(f)(g\sigma^{-1}b)\\
& = & G(f)(g)S(f)(\sigma^{-1})f(b)\\
& = & G(f)(g) \tau^{-1} f(b) \\
& = & \psi (G(f)(g) f(a)) = \psi(X_a(f)(ga))
\end{eqnarray*}

 In view of this, from now on I write $X(f)$ for $X_a(f)$, $a$ any basepoint. 
It follows easily that $X(f_2 \circ f_1) = X(f_2) \circ X(f_1)$ for any injections 
$f_1 : A \rightarrow B$ and $f_2 : B \rightarrow C$.}\footnote{In the sense that 
$X(f_2 \circ f_1)$ and $X(f_2) \circ X(f_1)$ are the same up to canonical isomorphisms.}

This allows us to define, for any 
tests $A' \in {\mathfrak G}(A)$, $B \in {\mathfrak G}(B)$, and any injection $f : A' \rightarrow B'$, a test-space morphism $X^{A}_{B}(f) : X(A) \rightarrow X(B)$ 
by
\[X^{A}_{B}(f) \ = \ hX(h^{-1}\circ f \circ g) g^{-1}\]
where $g \in G(A)$ with $gA = A'$ and $h \in G(B)$ with $hB = B'$ (and where, of course, inside the scope of $X$, 
$h^{-1}$ and $g$ represent, respectively, $g|_{A}$ and $h^{-1}|_{B'}$). The claim is that this is well-defined, i.e., independent of the 
particular choice of $g$ and $h$.

{\bf Proposition: 1} {\em Let $f : A' \rightarrow B'$ with $A' = gA = g'A$ and $B' = hB = h'B$, 
for $g, g' \in G(A)$ and $h, h' \in G(B)$. Then 
\[h X(h^{-1} \circ f \circ g ) g \ = \ h' X({h'}^{-1} \circ f \circ g') g'.\]}

{\em Proof:} We first show that if $f : A \rightarrow B$, $k \in G(A)_A$ and $\ell \in G(B)_B$, 
\[\ell X(\ell^{-1} \circ f \circ k) k^{1} = X(f).\]
Let $a \in A$ and $b \in B$. If $k = \sigma k'$ as in part (b) of Lemma 5 above, then 
$f(ka) = f(\sigma k' a) = f(\sigma a)$. Also note that, by part (a) of Lemma 5, 
for any $x \in A$, 
\begin{eqnarray*}
G(f)(k)f(x) = G(f) (\sigma k')f(x) & = & G(f)(\sigma)G(f)(k') f(x)\\
&  = & G(f)(\sigma)f(x) = S(f)(\sigma) f(x) = f(\sigma x).
\end{eqnarray*}
Now let $x = pa$, $p \in G(A)$, be any element of $X(A)$: 
\begin{eqnarray*}
X(f \circ k_A) k^{-1} (pa) & = & G(f \circ k_A)(k^{-1}p) f(ka)\\
&  = & G(f)(G(k)(k^{-1}p) f(\sigma a) \\
& = & G(f) (k (k^{-1} p) k^{-1}) f(\sigma a) \\
&  = & G(f)(p k^{-1}) f(\sigma a) \\
& = & G(f)(p) G(f)(k^{-1})f(\sigma a)\\
& = & G(f)(p)f(\sigma^{-1} \sigma a) = X(f)(pa).
\end{eqnarray*}
(Note the use of regularity in the third line.)
For $\ell \in G(B)_B$, again using regularity, we have
\begin{eqnarray*} 
\ell^{-1} X(\ell \circ f) (pa) = \ell^{-1} G(\ell \circ f)(p) \ell f(a) & = & \ell^{-1}G(\ell)(G(f)(p)) \ell f(a)\\
&  = & \ell^{-1} \ell G(f)(p) \ell^{-1} \ell f(a)\\
& =  & G(f)(p) f(a) = X(f)(pa).
\end{eqnarray*}
Combining these, we see that 
$\ell^{-1} X(\ell \circ f \circ k ) k^{-1} = X(f)$, as promised. 

Now let $f : A' \rightarrow B'$ with $A' = gA = g'A \in \G(A)$ and $B' = hB = h'B \in \G(B)$. Then 
$g' = gk$ and $h' = hj$ for $k \in G(A)_A$ and $\ell \in G(B)_B$, whence, 
\[ \ \ h' X({h'}^{-1} f g') {g'}^{-1}  =  h \ell X(\ell^{-1} h^{-1} f g k) k^{-1} g^{-1} = h X(h^{-1} f g) g^{-1}. \ \ \Box\]



Once we have that $X^{A}_{B}$ is well-defined, it follows that it behaves properly with respect to composition:

{\bf Lemma 6:} {\em
If $(G,j)$ is regular, then 
for all $A' \in \G(A)$, $B' \in \G(B)$ and $C' \in \G(C)$, 
and all injections $f_1 : A' \rightarrow B'$, $f_2 : B' \rightarrow C' \in {\mathfrak G}(C)$, 
\[X^{B}_{C}(f_2) \circ X^{A}_{B}(f_1) = X^{A}_{B}(f_2 \circ f_1).\]} 

{\em Proof:} Let $g \in G(A), h \in G(B), k \in G(C)$ with $gA = A', hB = B'$ and $kC = C'$, respectively; then we have  
\begin{eqnarray*} X^{B}_{C}(f_2) \circ X^{A}_{B}(f_1) & = & kX(k^{-1} \circ f_2 \circ h)h^{-1} hX(h^{-1} \circ f \circ g) \\
&  = & kX(k^{-1} \circ f_2 \circ h) \circ X(h^{-1} \circ f \circ g)g' \\ 
& = & k X(k^{-1} \circ f_2 \circ f_1 \circ g)g^{-1} = X^{A}_{B}(f_2 \circ f_1). \ \Box \end{eqnarray*}

{\em Notation:} Where $f : A' \rightarrow A''$ with $A', A'' \in {\mathfrak G}(A)$, I'll write $X_{A}(f)$ for $X^{A}_{A}(f)$.

{\blue {\bf Lemma 7:} {\em Let $(G,j)$ be regular. Then for all injections $f : A' \rightarrow B'$, where 
$A' \in \G(A)$  and $B' \in \G(B)$, and for all $g \in G(A)$, $h \in G(B)$, we have 
\[X^{A}_{B}(h \circ f \circ g) =  h \circ X^{A}_{B}(f) \circ g^{-1}.\]}

{\em Proof:} Let $B'' = h^{-1} B'$. Let $h' : B \rightarrow B'$ be any element of $G(B)$ with $h' B = B'$. 
Then we have $h h' A = B''$, whence, for any $g' \in G(A)$ with $g' A = A'$, we have  
\[X^{A}_{B}(h \circ f) = h h' X((h h')^{-1} \circ h \circ f \circ g ) g^{-1} = h h' X({h'}^{-1} \circ f \circ g) \circ g^{-1} = h X^{A}_{B}(f).\]
A similar argument shows that $X^{A}_{B}(f \circ g) = X^{A}_{B}(f) \circ g^{-1}$. $\Box$
} 

As a consequence, for all $g \in G(A)$, we have 
\[X^{A}_{A}(g|_{A'})x = gx\] 
for all $x \in X(A)$ and all $A' \in \G(A)$.  

{\blue Combining this observation with Lemma 6,} we see that, for a regular extension $(G,j)$, every $G-\Tesp$ morphism ${\mathfrak G}(A) \rightarrow {\mathfrak G}(B)$ has the form $X^{A}_{B}(f)$ for some $f : A' \rightarrow B'$, $A' \in {\mathfrak G}(A)$, $B' \in {\mathfrak G}(B)$. In fact, we can say a bit more.



{\red \subsection{A Canonical Form for $G$-$\Tesp$ Morphisms}

{\bf Lemma 8:} {\em  Let $G$ be regular. For any injection $f : A \rightarrow B$, and for all $g \in g(A)$, 
\[X(f)g = G(f)(g) X(f).\]}

{\em Proof:} Let $a \in A$. For any $x = pa \in X(A)$, $p \in G(A)$, we have 
\begin{eqnarray*}
\ \ X(f)(g x) = X(f)(gpa) & = & G(f)(gp)f(a)\\
 & = & G(f)(g)G(f)(p)f(a)\\
&  = & G(f)(g) X(f)(pa) = G(f)(g)X(f)(x). \ \Box
\end{eqnarray*}

This gives us 

{\bf Proposition 2:} {\em If $G$ is regular, then every $G$-$\Tesp$ morphism $\G(A) \rightarrow \G(B)$ has the form 
$g X(f)$ for some injection $f : A \rightarrow B$ and some $g \in G(B)$. }

{\em Proof:} 
A morphism ${\G}(A) \rightarrow {\G}(B)$ will have the form 
\[g_{n+1} X(f_n) \cdots g_2 X(f_1) g_1\]
where $f_{i} : A_{i} \rightarrow A_{i+1}$ are injections and $g_i \in G(A_{i})$ for $i = 1,...,n$. Applying 
Lemma 8 repeatedly gives us 
\[\begin{array}{l}
g_{n+1} X(f_n) g_{n} X(f_{n-1}) g_{n-1} \cdots g_2 X(f_1) g_1  \\ 
= \  g_{n+1} G(f_n)(g_n) G(f_n \circ f_{n-1})(g_{n-1}) \cdots G(f_n \circ \cdots \circ f_1)(g_1) X(f_n \circ \cdots \circ f_1). \ \Box 
\end{array}\]



It follows that $G$-$\Tesp$ morphisms are determined by their actions on 
a single test:

{\bf Corollary 2:} {\em Let $G$ be regular, and let $\phi, \phi' : X(A) \rightarrow X(B)$ be $G$-$\Tesp$ morphisms. If 
$\phi|_{A} = \phi'|_{A}$, then $\phi = \phi'$. }

{\em Proof:} Let $\phi = g X(f)$ and $\phi' = g' X(f')$ for $f, f' : A \rightarrow B$ and $g, g' \in G(B)$. Then 
for any $x \in A$, we have $\phi(x) = g X(f)(x) = g f(x)$ and $\phi'(x) = g X(f')(x) = gf'(x)$. So the injections 
$g \circ f : A \rightarrow X(B)$ and $g' \circ f' : A \rightarrow X(B)$ coincide. In particular, $gB = g'B =: B'$. 
Thus, regarding $g \circ f, g' \circ f'$ as injections $A \rightarrow B' \in \G(B)$, we have 
\[\ \ \phi = g X(f) = X^{A}_{B}(g \circ f) = X^{A}_{B}(g' \circ f') = g' X(f') = \phi'. \ \Box \]

This strengthens the analogy between $\GTesp$ and the category of finite-dimensional Hilbert spaces and 
unitary maps, with tests in $\G(A)$ playing much the same role in the former that orthonormal bases play in the latter.

{\bf Corollary 3:} {\em Every $\GTesp$ morphism $\phi : \G(A) \rightarrow \G(B)$ has the form 
$X^{A}_{B'}(f')$ for a unique $B' \in \G(B)$ and injection $f' : A \rightarrow B' \in \G(B)$. }

{\em Proof:} Let $\phi = g X(f)$ as in Proposition 2, where $f : A \rightarrow B$. Let 
$B' = gB \in \G(B)$ and let $f' = g \circ f$; then $X^{A}_{B'}(f')$ agrees with $\phi = g X(f)$ on $A$, 
and hence, by Corollary 2, $X^{A}_{B'}(f) = \phi$. Uniqueness of $B'$ and $f'$ is also immediate from Corollary 2. 
$\Box$

}

\section{Reasonable Extensions}

The functor $S : \Sinj \rightarrow \Grp$ has the very nice, and very reasonable, feature that if $A$ and $B$ are disjoint sets, 
then $S(A)$ and $S(B)$, as embedded in $S(A \cup B)$, are pairwise-commuting, in the sense that if $\sigma \in S(A)$ and $\tau \in S(B)$, 
then $\sigma \tau  = \tau \sigma$ in $G(A \cup B)$. 

{\bf Definition 8:} An extension $(G,j)$ of $S$ is {\em reasonable} iff,  for 
all disjoint sets $A$ and $B$, $G(A)$ and $G(B)$ commute pairwise in $G(A \cup B)$. 

Equivalently, $(G,j)$ is reasonable iff there exists a natural homomorphism $\phi : G(A) \times G(B) \rightarrow G(A \cup B)$ such that the diagram 
\[{\xymatrix@=12pt{  & G(A) \times G(B) \ar@{->}^{\phi}[dd]    & \\ 
G(A) \ar@{->}[ur] \ar@{->}[dr]  & &   G(B) \ar@{->}[ul] \ar@{->}[dl]  \\ &  G(A \cup B)  & }}\]
commutes (where the maps $G(A), G(B) \rightarrow G(A) \times G(B)$ are the canonical injections $a \mapsto (a,e)$ and $b \mapsto (e,b)$). 

The theories arising from reasonable extensions are particularly well-behaved, owing to the following

{\bf Lemma {\blue 9}:} {\em If $(G,j)$ is reasonable, then for any finite sets $A \subseteq B$, $G(A)$ fixes every point 
of $X(B \setminus A)$.}

{\em Proof:} Choosing a base-point $b \in B \setminus A$, we can model $X(B)$ as $G(B)/G(B \setminus b)$. As $G(A) \leq G(B \setminus b)$, we have $gb = b$ for every $g \in G(A)$. Now $X(B \setminus A)$ consists of 
points of $X(B)$ of the form $x = hb$ with $h \in G(B \setminus A)$. As $(G,j)$ is reasonable, if $g \in G(A)$ and 
$h \in G(B \setminus A)$, we have $ghb = hgb = hb$. $\Box$ 

Write $A \oplus B$ for the union of  {\em disjoint} sets $A$ and $B$. 
Recall from Section 2 that a test space ${\mathfrak A}$ is {\em algebraic} iff perspective events -- events having one common complementary event -- are complementary to exactly the same set of events.  That is, if 
$E \oplus F, F \oplus F'$ and $E' \oplus E$ are tests, then $E'$ and $F'$ are disjoint, and $E' \oplus F'$ is also a test.

{\bf Proposition 3:} {\em If $(G,j)$ is a reasonable {\red regular} extension, then  
\begin{mlist}
\item[(a)] ${\mathfrak G}(A)$ is algebraic for every $A$; 
\item[(b)] If $A \cap B = \emptyset$, then ${\mathfrak G}(A) = {\mathfrak G}(A \cup B)_{A}$ where $A$ is regarded as an event in ${\mathfrak G}(A \cup B)$. 
\item[(c)] If $f : A \rightarrow B$ is an injective mapping, then $X(f) : X(A) \rightarrow X(B)$ is a morphism of test spaces. 
\end{mlist}}

{\red {\em Proof:} 

(a) Let $B, C, B'$ and $C'$ be events of $\G(A)$ with $B \oplus C, B' \oplus B$ and $C \oplus C'$, tests in $\G(A)$. 
Without loss of generality, we can assume that $B \oplus C = A$. We must show that $B'$ and $C'$ are disjoint, 
and that $B' \oplus C'$ belongs to $\G(A)$.   Since $\G(A)$ is fully $G(A)$-symmetric, we can find an element 
$g \in G(A)$ fixing $C$ pointwise, and taking $B$ to $C'$. Since $(G,j)$ is regular, Lemma 4 implies that 
$g \in G(C)$. Similarly, we can find $h \in G(B)$ with $hC = B'$. Since $(G,j)$ is reasonable 
and $B \cap C = \emptyset$, $g$ and $h$ commute. We then have $ghA = gh(B \oplus C) = ghB \oplus hgC = gB \oplus hC = C' \oplus B'$. Since $ghA \in \G(A)$, we are done. 
 
(b) If $B \subseteq A$ and $B \sim C$ in $\G(A)$, then by (a), we have 
$C \co (A \setminus B)$. We can therefore find a bijection $g : A \rightarrow B \oplus C$ fixing $A \setminus B$ pointwise 
and sending $B$ to $C$. But then $g \in F^{A}_{A \setminus B} = G(B)$ by {\green Lemma 4}, whence $C \in \G(B)$. 

(c) It suffices to show that if $A' \in \G(A)$, then $X(f)(A) \sim X(f)(A')$ in $\G(B)$. Let 
$A' = gA$, so that $X(f)(A') = G(f)(g)(f(A))$. Let
$C = B \setminus f(A)$: then $G(f(A)) \leq G(B)$ fixes $C$ pointwise, by Lemma 9. As  
$G(f) : G(A) \rightarrow G(f(A))$  takes $g$ to $G(f)(g) \in G(f(A))$, we have 
\[G(f)(g)(B) = G(f)(g)(f(A)) \cup G(f)(g)(C) = X(f)(A') \cup C.\]
Since $X(f)(A') \cap C = G(f)(g)A \cap G(f)(g)C = \emptyset$, we have $X(f)(A') \co C \co f(A)$, whence, 
$X(f)(A') \sim f(A)$. $\Box$ 


{\red 
\section{Monoidal extensions}

Earlier, it was pointed out that $\G(A \times B)$ provides a natural candidate for a composite of the test spaces 
 $\G(A)$ and $\G(B)$. Ideally, one would like this to be a proper, non-signaling composite in the sense of Definition 2. 
 Beyond this, one would like the rule $\G(A), \G(B) \mapsto \G(A \times B)$ to be the object part of a symmetric monoidal structure on $G$-$\Tesp$, with $X : \FinInj \rightarrow \GTesp$ a monoidal functor. 

For finite sets $A$ and $B$, and for every $a \in A, b \in B$, let 
$\Phi^{b} : A \rightarrow A \times B$ and $\Phi_{a} : B \rightarrow A \times B$ be the 
mappings 
\[\Phi^{b} : x \mapsto (x,b) \ \ \mbox{and} \ \ \Phi_{a} : y \mapsto (a,y).\]
These give rise to homomorphisms 
\[\phi^b = G(\Phi^{b}) : G(A) \rightarrow G(A \times \{b\}) \leq G(A \times B)\]
and 
\[\phi_a = G(\Phi_a) : G(B) \rightarrow G(\{a\} \times B) \leq G(A\times  B)\]
If $G$ is reasonable, then $\phi^b(g)$ commutes with $\phi^{b'}(g)$ for 
$b \not = b'$ in $B$, by reasonability (as $A \times \{b\} \cap A \times \{b'\} = \emptyset$); similarly, 
$\phi_a(h)$ commutes with $\phi_{a'}(h)$ for $a \not = a' \in A$. We thus have a pair of 
canonical homomorphisms
\[G(A) \stackrel{\phi_1}{\longrightarrow} G(A \times B) \stackrel{\phi_2}{\longleftarrow} G(B).\]
given by  
\[\phi_1(g)  := \Pi_{b \in B} \phi^{b}(g) \ \ \ \mbox{and} \ \ \ \phi_2(h) := \Pi_{a \in A} \phi_{a}(h).\]

{\bf Definition 9:} A reasonable extension $(G,j)$ is {\em monoidal} iff, for all finite sets $A$ and $B$, $\phi_1(g)$ commutes with $\phi_2(h)$ for every $g \in G(A)$ and $h \in G(B)$. 

Thus, if $(G,j)$ is monoidal, we have a canonical homomorphism $\phi := (\phi_1 \otimes \phi_2) : G(A) \times G(B) \rightarrow G(A \times B)$.

{\bf Examples:}  (1) If $G = S$ and $\sigma \in S(A), \tau \in S(B)$, it is straightforward that 
$\phi_1(\sigma)(x,y) = (\sigma x, y)$ and 
$\phi^{2}(\tau)(x,y) = (x, \tau y)$. Thus, $S$ is monoidal. 
(2) Let $G = U$, the unitary extension discussed 
in Section 3.2. That is, $U(A) = U(\H(A))($, where $\H(A) = (\C^{A})^{\ast}$; regarding 
$A$ as an orthonormal basis for $\H(A)$, $\H(A \times B) = \H(A) \otimes \H(B)$. One can work out that 
$\phi_1 (u) = u \otimes \1$ for $u \in U(A)$ and $\phi_2(v) = \1 \otimes v$ for $v \in U(B)$. Thus, 
$\phi_1(u)$ and $\phi_2(v)$ commute, and so $U$ is monoidal. (3) Let $G$ be the ``graph" extension 
discussed in Section 3.2: $G(A) = S(A) \times S(A)$, 
with $j_A(\sigma) = (\sigma, \sigma)$. Then $X(A) = A \times A$; hence, 
$X(A \times B) = (A \times B) \times (A \times B)$. As discussed earlier, 
this is a reasonable extension. One can work out that $\phi_1(\sigma, \tau)((x,y),(x',y')) 
= ((\sigma x, y), (\tau x', y))$, while $\phi_2(\sigma,\tau)((x,y),(x',y')) = ((x,\sigma y), (x', \tau y'))$. 
Clearly, these commute, so the Graph extension is also monoidal. (The ``grid" example, which is not reasonable, is therefore also not monoidal.)

For the balance of this section, $(G,j)$ is a monoidal extension.

{\em Notation:} If $g \in G(A)$ and $h \in G(B)$, where $(G,j)$ is monoidal, let 
\[g \otimes h \ := \ \phi_1(g) \otimes \phi_2(h).\]

It is then straightforward that 
\begin{equation} (g \otimes h) (g' \otimes h') = gg' \otimes hh'\end{equation}
for all $g, g' \in G(A)$ and $h, h' \in G(B)$. It is also easy to check that 
if $\sigma \in S(A)$ and $\tau \in S(B)$, we have 
\[(\sigma \otimes \tau)(a,b) = (\sigma a, \tau b)\]
for all $a \in A, b \in B$. 

{\bf Lemma 10:}{\em Let $k \in K(A,a) = G(A \setminus a)$. Then $k \otimes e = \phi_1(k)$ fixes all points in $X(\{a\} \times B)$; 
similarly, if $\ell \in K(B,b)$, then $e \otimes \ell = \phi_2(\ell)$ fixes all points of $X(A \times \{b\})$. } 

{\em Proof:} Since $\Phi^{b}(A \setminus a) = (A \setminus a) \times \{b\} \subseteq (A \setminus A) \times B$, 
reasonableness of the extension tells us that $\phi^{b}(k) = G(\Phi^{b})(k) \in G((A \setminus \{a\}) \times \{b\}) \leq 
G((A \setminus a) \times B)$ fixes 
every point in $X(A \times B \setminus ((A \setminus \{a\}) \times B))$, i.e, in 
$X(\{a\} \times B)$. As this holds for all $b \in B$, $\phi_1(k) = \Pi^{b \in B} \phi^{b}(k)$ also 
fixes all points in $X(a \times B)$. The second claim is proved in the same way. $\Box$ 

A consequence is that if $k \in K(A,a)$ and $\ell \in K(B,b)$, then 
$k \otimes \ell$ fixes $(a,b) \in X(A) \times X(B)$, i.e., $k \otimes \ell \in K(A \times B, (a,b))$. This gives us 
a well-defined mapping 
\[X(A) \times X(B) \ \stackrel{\otimes}{\longrightarrow} \ X(A \times B)\]
taking any $x = ga \in X(A)$ and $y = hb \in X(B)$ to 
\[x \otimes y \ := \ (g \otimes h)(a,b).\]
We then have, for any $g'\in G(A)$ and $h' \in G(B)$, that 
\[(g' \otimes h')(x \otimes y) = (g' \otimes h')(g \otimes h)(a,b) = (g'g \otimes h'h)(a,b) = g'ga \otimes h'hb = g'x \otimes h'y.\]
It follows from this that $x \otimes y$ is independent of the choice of basepoints $a \in A, b \in B$: if 
$x = ga = g'a'$ and $y = hb = h'b'$, with $a' \in A, b' \in B$, then let $a' = \sigma a$ and $b' = \tau b$ 
for $\sigma \in S(A)$ and $\tau \in S(B)$. We have $ga = g'\sigma a$, whence, $g = g'\sigma k$ where $ka = a$, 
and similarly $h = h'\tau \ell$ with $\ell b = b$. Thus, 
\begin{eqnarray*}
(g' \otimes h')(a',b') = (g' \otimes h')(\sigma k a, \tau \ell b) & = & (g' \otimes h')(\sigma k \otimes \tau \ell)(a,b)\\ 
& = & (g' \sigma k \otimes h' \tau \ell)(a,b) = (g \otimes h)(a,b).
\end{eqnarray*}



\subsection{Composites in $G$-$\Tesp$}

We are now in a position to establish that states on  $\G(A \otimes B) := \G(A \times B)$ are non-signaling:

{\bf Theorem 1:} {\em Let $G$ be monoidal and reasonable. The mapping 
\[\otimes : X(A) \times X(B) \rightarrow X(A \times B)\] 
defined above makes $\G(A \times B)$ a  composite of $\G(A)$ and $\G(B)$, in the sense  of Definition 2.}

{\em Proof:} It will suffice to show that the image under $\otimes$ of any test in the Foulis-Randall 
product $\G(A)\G(B)$ 
belongs to $\G(A \times B)$. Let $B_x \in \G(B)$ for each $x \in A$. Then $B_x = h_x(B)$ for some $h_x \in G(B)$. 
Letting $\hat{h}_x = \phi_x(h_x) \in G(\{x\} \times B) \leq G(A \times B)$, we have  
$\hat{h}_x (x \otimes y) = x \otimes h_x y$ 
for all $y \in B$, while $\hat{h}_x (x' \otimes y) = x' \otimes y$ for all 
$y \in B$ and all $x' \in A$ with $x' \not = x$. Now let $\hat{h} = \Pi_{x \in A} \hat{h}_x$: then 
$\hat{h}(x,y) = \hat{h}_x(x, y) = x \otimes h_x y$ for all $x, y \in A \times B$, whence, 
\[\bigcup_{x \in A} x \otimes B_x = \bigcup_{x \in A} x \otimes \hat{h}_{x}(B) = 
\bigcup_{x \in A} \hat{h}(\{x\} \times B) = \hat{h}(A \times B) \in \G(A \times B).\]
Now let $A' = g A \in \G(A)$. Given a test $B_z \in \G(B)$ for each $z \in A'$, re-index so that $B_x := B_z$ with $z = gx$ 
for each $x \in A$. 
Then 
\[\bigcup_{z \in A'} z \otimes B_{z} = \bigcup_{x \in A} gx \otimes B_x = \bigcup_{x \in A} \phi_1(g)(x \otimes B_x) 
= \phi_1(g) (\bigcup_{x \in A} x \otimes B_x) \in \G(A \times B).\]
Thus, $\G(A \times B)$ contains the image of $\stackrel{\longrightarrow}{\G(A)\G(B)}$ under $\otimes$. The same argument, using the fact that $\phi_2(h)(g_y x \otimes y) = g_y x \otimes hy$, shows that 
it also contains the image of $\stackrel{\longleftarrow}{\G(A)\G(B)}$. $\Box$ 

{\em Remark:} This {\em almost} works without monoidality. If $(G,j)$ is regular and reasonable, we can define 
{\em two} mappings $\otimes_{L}, \otimes_{R} : X(A) \times X(B) \rightarrow X(A \times B)$, given, respectively, by 
\[ga \otimes_{L} hb = \phi_2(h)\phi_1(g)(a, b)  \ \ \mbox{and} \ \ ga \otimes_{R} hb = \phi_1(g)\phi_2 (h)(a,b)\]
Then $\G(A \times B)$ contains the image of the forward product $\stackrel{\longrightarrow}{\G(A)\G(B)}$ 
under $\otimes_{R}$, and the image of the backward product under $\otimes_L$. 

\subsection{Monoidality of $G$-$\Tesp$} 

We now show that if the extension $(G,j)$ is both monoidal and regular, the composition rule $\G(A), \G(B) \mapsto \G(A \otimes B)$ is the object part of a symmetric monoidal structure on $G$-$\Tesp$. 

We begin with some preliminary observations. 
Suppose
that $x \in X(A)$ and $y \in X(B)$ with $x = g'a$ and $y = h'b$. By definition, 
\[x \otimes y = (g' \otimes h')(a,b) = (g' \otimes h')(a \otimes b).\]
Hence, for any $g \in G(A)$ and $h \in G(B)$, we have
\[(g \otimes h)(x \otimes y) = (g \otimes h)(g' \otimes h')(a,b) = (gg' \otimes hh)(a,b) = gx \otimes hy.\]

{\bf Lemma 11:} {\em The mapping 
\[g X(f_1), h X(f_2) \mapsto (g \otimes h) X(f_1 \times f_2)\]
is well-defined, where $f_i : A_i \rightarrow B_i$ for $i = 1,2$, $g \in G(B_1)$ and $h \in G(B_2)$. }

{\em Proof:} Let $g X(f_1) = g' X(f_{1}')$ and $h X(f_2) = h' X(f_{2}')$. Then $g f_1 = 
g'  f_{1}'$ on $A$ and $h f_2 = h' f_{2}'$ on $B$; hence, $ g f_1 \times h f_2 = g' f_{1}' \times h' f_{2}'$ on $A \times B$, whence, 
\[(g \otimes h) X (f_1 \times f_2) = X^{AB}_{CD}((g \otimes h) f_1 \times f_2) =  
X^{AB}_{CD} (g' f_{1}' \times h' f_{2}') = (g' \otimes h')X (f_{1}' \times f_{2}').\]
Similarly, if $h X(f_2) = h'X(f_{2}')$, then 
\[(e \otimes h) X^{AB} (f_1 \times f_{2}') = (e \otimes h')X(f_1 \times f_{2}'). \Box\]

We can now define 
\[g X(f_1) \otimes h X(f_2) = (g \otimes h) X(f_1 \times f_2).\]
In view of Proposition 2, this gives us, for any morphisms $\phi_i : \G(A_i) \rightarrow \G(B_i)$, a 
canonical morphism 
\[\phi_1 \otimes \phi_2 : \G(A_1) \otimes \G(A_2) \rightarrow \G(B_1) \otimes \G(B_2).\] 

Our goal now is to show that this plays nicely with products of elements of $X(A_1)$ and $X(A_2)$. 

{\bf Notation:} If $M \subseteq A$, $N \subseteq B$, let 
\[\phi^{N}(g) = \Pi_{b \in N} \phi^{b}(g) \ \mbox{and} \  \phi_{M}(h) = \Pi_{a \in M} \phi_{a}(h).\]

{\bf Lemma 12:} {\em Let $a \in A$, $b \in B$. Then for all $h \in G(N) \leq G(B)$,  
\[ ga \otimes h b = \phi^{N}(g)\phi_{M}(h)(a,b).\]}

{\em Proof:}  We have 
\[g a \otimes h b = \phi_1(g) \phi_2(h) (a,b) = \phi_1(g) \phi_a(h)(a,b).\]
Let $z := \phi_{a} (h) (a,b)$. Note that this lies in $G(aN)(a,b) \subseteq X(aN) \subseteq X(AB)$. For 
$b' \in B \setminus N$, $\phi^{b'}(g) \in G(Ab')$ fixes points of $X(AB \setminus Ab')$ 
As $N \subseteq B \setminus b'$, $aN \subseteq A(B \setminus b')$, so $\phi^{b'}(g)$ fixes points of 
$X(aN)$ pointwise. In particular, $\phi^{b'}(g)$ fixes $z$ for every $b' \in B \setminus N$. 
Thus, 
\[ga \otimes hb = \phi_1 (g) z  = \Pi_{b' \in B} \phi^{b' \in B}(g) z = \Pi_{b' \in N} \phi^{b} z = \phi^{N}(g) z.\]
Finally, note that since $a \in M$, $\phi_2(h)(a,b) = \phi_{a}(h) (a,b) = \Pi_{a' \in M} \phi_{a'}(h) (a,b)$. $\Box$ 

It follows that $x \otimes y$ as defined in $X(M \times N)$ coincides with $x \otimes y$ as defined in $X(A \times B)$. 

{\bf Lemma  13:} {\em Let $f_1 : A_1 \rightarrow B_2$ and $f_2 : A_2 \rightarrow B_2$ be injections, $i = 1,2$. Then for all $g_i \in G(A_i)$, 
\[G(f_1 \times f_2)(g_1 \otimes g_2) = G(f_1)(g_1) \otimes G(f_2)(g_2).\]} 

{\em Proof:} It is enough, in view of Corollary 2, to show that 
\[G(f_1 \times f_2)(g_1 \otimes g_2)(b_1, b_2) = G(f_1)(g_1)b_1 \otimes G(f_2)(g_2)b_2\]
for all $b_1, b_2 \in B_1 \times B_2$. 

For $a \in A_1$ and $b \in A_2$, let $\Phi_a : A_2 \rightarrow A_1 \times A_2$ and $\Phi^b : A_1 \rightarrow A_1 \times A_2$ be the mappings 
$\Phi_a(y) = (a,y)$ and $\Phi^b (x) = (x,b)$; also, for $c \in B_1$, let $\Psi^{c} : B_1 \rightarrow B_1 \times B_2$ 
be the mapping $\Psi^{c}(z) = (z,c)$. Then we have
\[(f_1 \times f_2) \circ \Phi^{b}  = \Psi^{f_2(b)} \circ f_1:\]
Then 
\[G(f_1 \times f_2)(g_1 \otimes g_2) = G(f_1 \times f_2)\phi_1(g_1) \phi_2(g_2).\]
Now, $\phi_1(g_1) = \Pi^{b \in A_2} \phi^{b}(g_1) = \Pi^{b \in B_2} G(\Phi^{b})(g_1)$. 
We have 
\begin{eqnarray*}
G(f_1 \times f_2)(\phi^{b}(g_1)) & = & G(f_1 \times f_2)(G(\Phi^b)(g_1)) \\
& = & G((f_1 \times f_2) \circ \Phi^{b})(g_1) \\
& = & G( \Phi^{f_2(b)} \circ f_1)(g_1) = \phi^{f_2(b)}(G(f_1)(g_1))
\end{eqnarray*}
In particular, 
\[\Pi^{b \in B_2} G(f_1 \times f_2)(\phi^{b}(g_1)) = \Pi^{c \in f_2(B_2)} \phi^{c}(G(f_1)(g_1)).\]
If we let $N = F_2(B_2)$, this is, more briefly, 
\[G(f_1 \times f_2)(\phi_1(g_1)) = \phi^{N}(G(f_1)(g_1)).\]
In the same way, we have 
\[G(f_1 \times f_2)(\phi_2(g_2)) = \phi_{M}(G(f_2)(g_2)).\]
Applying this to $(b_1, b_2) = (f_1(a_1), f_2(a_2))$, we have 
\[G(f_1 \times f_2)(\phi_1(g_1) \phi_2(g_2))(b_1, b_2) = \phi^{N}(G(f_2)(g_2))\phi_{M}(G(f_1)(g_1))(b_1,b_2).\]
By Lemma 12 above, this last equals $G(f_1)(g_1)b_1 \otimes G(f_2)(g_2)b_2$. $\Box$ 

{\bf Lemma 14:} {\em With notation as above, 
\[X(f_1 \times f_2)(x \otimes y) = X(f_1)x \otimes X(f_2)y\]for all $x \in X(A_1)$ and $y \in X(A_2)$. } 

{\em Proof:} 
\begin{eqnarray*}
X(f_1 \times f_2)(x \otimes y) & = & X(f_1 \times f_2)(g_1 \otimes g_2)(a,b) \\
& = & G(f_1 \times f_2)(g_1 \otimes g_2)(f_1(a), f_2(b))\\
& = & (G(f_1)(g_1) \otimes G(f_2)(g_2))(f_1(a), f_2(b))\\
& = & (G(f_1)(g_1) \otimes G(f_2)(g_2))(f_1(a) \otimes f_2(b))\\
& = & G(f_1)(g_1)f_1(a) \otimes G(f_2)(g_2)(f_2(b))\\
& = & X(f_1)(g_1 a) \otimes X(f_2)(g_2 b)
\end{eqnarray*}
where, in the second step, we are appealing to Lemma 13. $\Box$ 

It now follows that for arbitrary morphisms $\phi_1 = g_1 X(f_1) : \G(A_1  \rightarrow \G(B_1)$ and 
$\phi_2  = g_2 X(f_2) : \G(A_2) \rightarrow \G(B_2)$, we have 
\[(\phi_1 \otimes \phi_2)(x \otimes y) = \phi_1(x) \otimes \phi_2(y)\]
for all $x \in X(A_1)$ and $y \in X(A_2)$. Given morphisms $\psi_i : \G(B_i) \rightarrow \G(C_i)$, $i = 1,2$, 
we see that $(\psi_1 \otimes \psi_2) \circ (\phi_1 \otimes \phi_2)$ and 
$(\psi_1 \circ \phi_1) \otimes (\psi_1 \otimes \phi_2)$ agree on all elements of $X(A_1 \times A_2)$ 
of the form $x \otimes y$. In particular, they agree on $A_1 \times A_2$. 

We have shown that $\otimes$ is bifunctorial on $\GTesp$. Since functor $X : \FinInj \rightarrow \GTesp$ 
converts $\times$ to $\otimes$, it can be used to transfer the symmetric monoidal structure on the former 
to a symmetric monoidal structure on the latter. That is, if 
\[\alpha_{A,B,C} : A \times (B \times C) \rightarrow (A \times B) \times C)\]
\[\sigma_{A,B} : A \times B \rightarrow B \times C\]
\[\lambda_{A} : 1 \times A \rightarrow A; \ \rho_{A} : A \times 1 \rightarrow A\]
are the components of the natural transformations $\alpha, \sigma, \lambda, \rho$ defining the standard symmetric monoidal structure on $\FinInj$, where 
$1 = \{0\}$ is the tensor unit in $\FinInj$, then 
we would like to show that 
\[X(\alpha_{A,B,C}) : \G(A) \otimes (\G(B) \otimes \G(C)) \rightarrow (\G(A) \otimes \G(B)) \otimes \G(C)\]
\[X(\sigma_{A,B}) : \G(A) \otimes \G(B) \rightarrow \G(B) \otimes \G(B)\]
\[X(\lambda_{A}) : \G(1) \otimes \G(A) \rightarrow \G(A); \ G(\rho_A) : \G(A) \otimes \G(1) \rightarrow \G(A)\]
define natural transformations $X(\alpha)$, $X(\sigma), X(\lambda)$ and $X(\rho)$, and give a symmetric monoidal structure on $\GTesp$. The necessary coherence conditions will lift directly from those 
for $\alpha, \sigma, \lambda$ and $\rho$. What is needed, then, is to show that $X(\alpha)$, $X(\sigma), X(\lambda)$ 
and $X(\rho)$ are indeed natural in $G-\Tesp$. Since naturality for morphisms of the form $X(f)$, $f : A \rightarrow B$ follows immediately from the functoriality of $X$ and the definition of $\otimes$, what remains is to 
establish naturality for morphisms $ g : \G(A) \rightarrow \G(A)$ arising from elements $g \in G(A)$. 

{\bf Lemma 15:} {\em Let $\alpha = \alpha_{A,B,C}$. Then 
$G(\alpha)(g \otimes (h \otimes \ell)) = (g \otimes h) \otimes \ell$ for 
all $g \in G(A), h \in G(B)$ and $\ell \in G(C)$. }

{\em Proof:} The proof is somewhat tedious, so I will provide only a sketch. Let $A, B$ and $C$ be finite sets 
and let $\alpha = \alpha_{A,B,C} : (a,(b,c)) \mapsto ((a,b),c)$ for $a \in A, b \in B$ and $c \in C$. Let 
$e_1, e_2$ and $e_3$ be the identity elements of $G(A), G(B)$ and $G(C)$, respectively.  
We have canonical mappings 
\[\phi^{1,23}_1 : G(A) \rightarrow G(A) \otimes (G(B) \otimes G(C))\]
\[\phi^{1,2}_1 : G(A) \rightarrow G(A) \otimes G(B), \ \ \mbox{and} \ \ \phi^{12,3}_1 : G(A) \otimes G(B) \rightarrow (G(A) \otimes G(B)) \otimes G(C).\]
For $g \in G(A)$, we have 
\[g \otimes (e_1 \otimes e_3) = \phi^{1,23}_{1}(g) \ \ \mbox{and} \ \ (g \otimes e_1) \otimes e_2 = \phi^{12,3}(\phi^{1,2}_{1} (g)).\]
I claim that 
\begin{equation} 
G(\alpha)(g \otimes (e_2 \otimes e_3)) = (g_1 \otimes e_2) \otimes e_3;
\end{equation}
equivalently, that 
\[G(\alpha) \circ \phi^{1,23}_{1} = \phi^{12,3} \circ \phi^{1,2}_{1}.\]
For $(b,c) \in B \times C$,  let 
\[\Phi^{bc} : A \rightarrow A \times (B \times C)\]
be the mapping $\Phi^{bc}(a) = (a,(b,c))$, and let $\phi^{bc} : G(A) \rightarrow G(A \times (B \times C))$ be 
given by $\phi^{bc}(g) = G(\Phi^{bc})(g)$. Then we have 
\[\phi^{1,23}_{1} : G(A) \rightarrow G(A \times (B \times C))\] 
given by 
\[\phi^{1,23}_{1}(g) = \Pi_{b \in B, c \in C} \phi^{bc}(g).\]
Note that 
\[(\alpha \circ \Phi^{bc})(a) = \alpha((a,(b,c)) = ((a,b),c)) = \Phi_{12,3}^{c}(\Phi_{1,2}^{b}(a)),\]
where $\Phi_{1,2}^{b} : A \rightarrow A \times B$ and $\Phi_{12,3}^{c} : (A \times B) \rightarrow (A \times B) \times C$ 
are the mappings $a \mapsto (a,b)$ and $(a,b) \mapsto ((a,b),c)$, respectively. Defining 
$\phi^{b} = G(\Phi_{1,2}^{b})$ and $\phi^{c} = G(\Phi_{12,3}^{c})$, we have 
\[\phi^{1,2}_{1} = \Pi_{b \in B} \phi^{b}\ \ \mbox{and} \ \ \phi^{12,3} = \Pi_{c \in C} G(\Phi^{c}).\]
Hence, 
\begin{eqnarray*} 
G(\alpha)(\phi^{1,23}_{1}(g)) & = & \Pi_{b \in B, c \in C} G(\alpha \circ \Phi^{bc}(g)) \\
& = & \Pi_{c \in C, b \in B} G(\Phi^{c} \circ \Phi^{b})(g) \\
& = & \Pi_{c \in C, b \in B}  G(\Phi^{c})(G(\Phi^{b}(g))) \\
& = & \Pi^{c \in C} G(\Phi^{c})(\Pi_{b \in B} G(\Phi^{b}(g)))\\
& = & \phi^{12,3}_{1} (\phi^{1,2}_{1}(g)) 
\end{eqnarray*}
In a similar way, one shows that, for $h \in G(B)$ and $\ell \in G(C)$, 
\begin{equation} G(\alpha)(e_A \otimes (h \otimes e_C)) = (e_A \otimes h) \otimes e_C \end{equation} 
and
\begin{equation} 
G(\alpha)(e_A \otimes (e_B \otimes \ell)) = (e_A \otimes e_B) \otimes e_{\ell}.
\end{equation} 
Combining the three identities (5), (6) and (7), and using the fact that $G(\alpha)$ is a homomorphism, together 
with the identity (3) above (or the 
bifunctoriality of $\otimes$) gives the desired result. $\Box$

This gives us at once that, if $x = ga, y = hb$ and $z = \ell c$ for base points $a \in A, b \in B, c \in C$, 
\begin{eqnarray*}
X(\alpha)(x \otimes (y \otimes z)) & = & G(g \otimes (h \otimes \ell))\alpha(x,(y,z)) \\
& = & ((g \otimes h) \otimes \ell)((x,y),z) 
= (x \otimes y) \otimes z. \end{eqnarray*}
Using this, it is easy to verify that $X(\alpha)$ is natural in $\GTesp$ (use Corollary 3).
It is straightforward that 
\[X(\sigma)(x \otimes y) = y \otimes x\]
and 
\[X(\lambda_{A})(0 \otimes x) = x, \ \ \mbox{and} \ \ X(\rho_A)(x \otimes 0) = x.\]
This, in turn, gives us the naturality of $X(\lambda)$ and $X(\rho)$ in $\GTesp$.

In summary, we have 

{\bf Theorem 2:} {\em If $(G,j)$ is regular and monoidal, then $\GTesp$ carries a canonical 
symmetric-monoidal structure, and $X : \FinInj \rightarrow \GTesp$ is a monoidal functor.}

\section{Conclusions and Directions for Further Work} 

These results obtained here suggest many interesting problems for further study, of which I will mention three.

1) First, one would like to find {\em categorical} conditions on an extension $(G,j)$ that are sufficient to make ${\mathfrak G}(A \times B)$ a genuine composite in the sense of Section 2. In view 
of Theorem 1, if $(G,j)$ is both regular and monoidal, ${\mathfrak G}(A) \times {\mathfrak G}(B)$ is canonically embedded in ${\mathfrak G}(A \times B)$, and states on the latter restrict to non-signaling states on the former. What is required, then, is that every product state on ${\mathfrak G}(A) \times {\mathfrak G}(B)$ extend to a product state on ${\mathfrak G}(A \times B)$. 
 
There is, of course, the danger that this condition could be satisfied trivially, i.e., that $\Omega({\mathfrak G}(A))$ be empty for all $A$. In order for the theory associated with $(G,j)$ to be of real interest, it is important that ${\mathfrak G}(A)$ host a rich state space. A test space ${\mathfrak A}$ is {\em sharp} iff, for every outcome $x \in X({\mathfrak A})$, there is a unique state $\epsilon_x \in \Omega({\mathfrak A})$ with $\alpha(x) = 1$. Call an extension $(G,j)$ sharp iff, for every finite set $A$, the test space ${\mathfrak G}(A)$ is sharp. If we assume both that  $(G,j)$ is sharp {\em and} that ${\mathfrak G}(A \times B)$ is a product for all $A$ and $B$, and, finally, that 
the state spaces of the factors are finite-dimensional, 
then one can show (as outlined in \cite{Wil09b}) 
that, for such an extension, the category $G-\Tesp$ satisfies most of Hardy's axioms \cite{Har01} for finite-dimensional quantum mechanics. 

(2) In a different direction, in the discussion of section 4 one would like to replace the rather impoverished category $\Sinj$ of finite sets an injective mappings by a richer category, such as the category $\Set$ of finite sets and mappings or the category $\Rel$ of finite sets and relations. One can do this by replacing the category $\Grp$ of groups and homomorphisms, by the category $\Grel$ of groups and polymorphisms (that is, subgroups of product groups, regarded as relations). If  
$f : A \rightarrow B$ is any mapping between sets $A$ and $B$, define \[S(f) = \{ (\sigma, \tau) \in S(A) \times S(B) | f \sigma = \tau f \} \leq S(A) \times S(B):\]
then $S(g \circ f) \subseteq S(g)S(f)$ (here reversing the usual order of relational multiplication), so we can regard $S$ as a lax functor from $\SET$ to $\Grel$. One can similarly regard $S$ as a functor $\Rel \rightarrow \Grel$. 

(3) Finally, there is the following piece of unfinished technical business. At present, I know of no reasonable, regular extension that is not monoidal. It would be of interest (to me, at any rate) to have such an example, and even more interesting to show that none exists.

\end{document}